\documentclass[sigconf, screen, authorversion, nonacm=true]{acmart}

\usepackage{amsmath}
\usepackage{graphicx}
\usepackage{hyphenat}
\usepackage[utf8]{inputenc}
\usepackage{tabularx}
\usepackage{multicol}
\usepackage{xcolor}
\usepackage{tikz}
    \usetikzlibrary{positioning, through, arrows, arrows.meta, external}

\usepackage{caption}
\usepackage{subcaption}
\usepackage{mdframed}

\definecolor{new_stuff}{RGB}{46, 204, 113}

\begin{document}

\title{On deceiving malware classification with section injection}


\author{Adeilson Antônio da Silva}
\orcid{0000-0002-9851-8910}
\email{adeilson.antonio@ufba.br}
\affiliation{%
  \institution{Federal University of Bahia}
  \country{Brazil}
}

\author{Mauricio Pamplona Segundo}
\orcid{0000-0003-4529-5757}
\email{mauriciop@mail.usf.edu}
\affiliation{%
  \institution{University of South Florida}
  \country{US}
}

\markboth{IEEE Signal Processing Letters,~Vol.~14, No.~8, February~2020}%
{Shell \MakeLowercase{\textit{et al.}}: Bare Demo of IEEEtran.cls for IEEE Journals}

\begin{abstract}
We investigate how to modify executable files to deceive malware classification systems. This work's main contribution is a methodology to inject bytes across a malware file randomly and use it both as an attack to decrease classification accuracy but also as a defensive method, augmenting the data available for training. It respects the operating system file format to make sure the malware will still execute after our injection and will not change its behavior. We reproduced five state-of-the-art malware classification approaches to evaluate our injection scheme: one based on GIST+KNN, three CNN variations and one Gated CNN. We performed our experiments on a public dataset with 9,339 malware samples from 25 different families. Our results show that a mere increase of 7\% in the malware size causes an accuracy drop between 25\% and 40\% for malware family classification. They show that a automatic malware classification system may not be as trustworthy as initially reported in the literature. We also evaluate using modified malwares alongside the original ones to increase networks robustness against mentioned attacks. Results show that a combination of reordering malware sections and injecting random data can improve overall performance of the classification.  Code available at \url{https://github.com/adeilsonsilva/malware-injection}.
\end{abstract}



\settopmatter{printfolios=true,printacmref=false}
\maketitle

\section{Introduction}
\label{sec:introduction}

Malware - a short term for malicious software - is described by Sikorski \textit{et al.} \cite{Sikorski2012} by their action:

\begin{quote}
Any software that does something that causes harm to a user, computer, or network can be considered malware [\ldots]
\end{quote}

These applications, purposedly built with intentions of reading, copying, or modifying information from computer systems - often without user consent - pose a high threat for modern information systems~\cite{Li2020,Microsoft2019,Symantec2019}. The early detection of such malware is vital to minimize their effects on an organization, or even among regular users.

In this work we discuss strategies related to the \textbf{classification} (\textit{i. e.} which kind of malware is it?) of malware samples using only their raw bytes as inputs to machine learning algorithms. These strategies can be seen as part of the static analysis of samples, a very important stage in a malware detection pipeline, in which is necessary to provide the classification without executing the file being analyzed. It is important to stress that these methodologies are not to be used as the solely strategy to detect malware samples, but as the first one in a multi-step chain of procedures. Despite that, due to their fast execution times and lack of human interaction, they are still an integral part of such a pipeline \cite{M3652020, Chen2020}.

We present here a simple way to modify a software file to deceive systems built to classify malware examples into families. Our method builds upon the idea of injecting bytes into the executable file \cite{Anderson2017}. We seek to insert bytes in different parts of a malware. By doing so, we aim to deceive malware classifiers and preserve the original functionality while hindering the detection of injected data. To accomplish that, we create rules of injection that respect the file format of the operating system the malware will infect. We can not only define how many bytes we inject but also how they spread over the file. More importantly, we explore two approaches: 
\begin{description}
    \item[Random injection:] inserting random bytes, so that we do not require any knowledge about the systems to be deceived
    \item[Adversarial injection:] inserting bytes taken from families different from the sample being evaluated
\end{description}

The classification approaches evaluated in this work are based on methods that learn straight from the raw bytes of the file, ranging from methodologies that reinterpret the sample as an grayscale image up to preprocessing each sample as a 1D vector in their execution \cite{Nataraj2011, Grosse2016, Athiwaratkun2017, Yue2017, Chen2018, Chen2020, Raff2017, Le2018, Su2018, Khormali2019, Benkraouda2021}. We want to evaluate the vulnerability of these variants to the already known adversarial examples \cite{Goodfellow2015}, an approach with increasing popularity in the literature, especially in the context of malware\cite{Grosse2016,Anderson2017,Al-Dujaili2018,Khormali2019,Demetrio2019,Demetrio2021,Benkraouda2021,Lucas2021}. There are some limitations that must be observed, tough, since the pertubations added to malware samples must be drawn from a discrete domain. It differs from other types of data, such as images. Also, executable files have strict standards, which means byte ordering is relevant in some parts of the file. As mentioned earlier, we tried to conform our manipulations to the expected stantards in order to preserve the functionality of the malware samples.

The rest of this paper is presented as follows: in Section \ref{sec:related-works} we compare our methodologies to other present in the literature. In Section \ref{sec:data-injection} we present how we generate and add data between sections of a PE file. Section \ref{sec:malware-classification} discusses the machine learning algorithms evaluated in this paper for malware classification. In Section \ref{sec:results} we discuss our evaluation strategies and their results, finishing with our conclusions in Section \ref{sec:conclusion}.

Our contributions can be summarized as follows:
\begin{enumerate}
    \item We provide a framework to inject data into PE files that leverages all the alignments required to preserve its functionality. It can inject any sort of data (either random or from a different file) in multiple positions of the file, not only at the end (padding).
    \item We evaluate how different deep neural networks architectures proposed for malware classification behave in multiclass classification scenarios. We want to assess the difficulties behind separating a given sample from other samples from the same kind.
    \item We evaluate how the aforementioned networks behave when dealing with injected samples. Our goal here is to assess how our attacks impact the classification of these networks, in regards both of the location and also the amount of injected data.
\end{enumerate}

\section{Background}
\label{sec:background}

Modern operating systems, such as Linux and Microsoft Windows, use the concept of sections to read an executable file, load it to the memory, and run its instructions. It is necessary to know and follow the file format specifications to be able to insert data in different parts of an executable and preserve its functionality. Since there are differences between the files accepted by each operating system, designing a system-agnostic injection scheme is impracticable. For this reason, we focus on the Microsoft Windows' PE32 format, as it is the only one included in publicly available malware datasets~\cite{Nataraj2011,Ronen2018}.

\begin{figure}[!ht]
    \begin{center}
        \includegraphics[width=.4\linewidth]{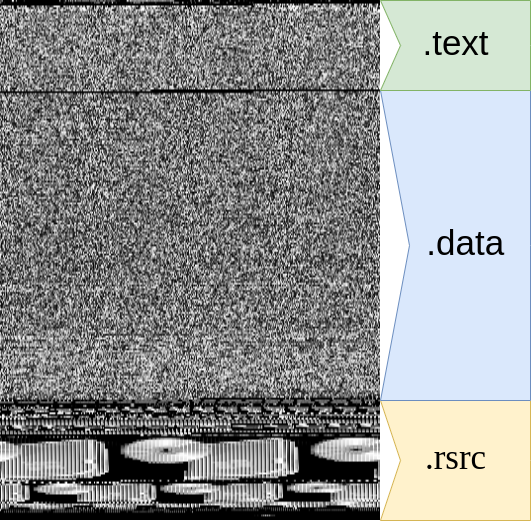}
        \caption{Illustration of the sections of a PE32 executable file.}
        \label{executable-sections}
    \end{center}
\end{figure}

As shown in Figure~\ref{executable-sections}, PE32 sections provide information about the executable, such as its instructions (``.text''), its variables(``.data''), and resources it uses (``.rsrc''). Following this layout, our strategy consists of injecting non-executable sections like ``.data'' to the file. This way, the set of instructions does not change, and the only way to decide whether an injected section is in use or not is through execution.

\section{Related Works}
\label{sec:related-works}

We start by presenting previous works which performed malware classification. We also discuss other methodologies for malware perturbation and what authors have done to minify these effects.


\subsection{Malware classification}
\label{sub:rw-classification}

Nataraj~\emph{et~al.}~\cite{Nataraj2011} presented a method to transform software into images and classify them according to their malware family. In this context, a family is a set of software files with high similarity of instructions and behavior. Follow up works explored this idea using different feature extraction (\emph{e.g.}, GIST, Local Binary Patterns, Scale-Invariant Feature Transform) and classification (\emph{e.g.}, Support Vector Machines, K-Nearest Neighbours (KNN)) methods~\cite{Agarap2018,Liu2019}.

The growth in Deep Learning research led to the exploration of neural networks for malware classification. Recent works applied different architectures for this task, either by extracting features from the file (\emph{e.g.}, system calls, imported libraries, functions in use)~\cite{Pascanu2015,Saxe2015,Athiwaratkun2017,Anderson2018} or by using the raw bytes from the data as input~\cite{Raff2017,Raff2018,Haddadpajouh2018}. Some of them achieve high classification accuracy by training Convolutional Neural Networks (CNNs) from scratch~\cite{Su2018,Liu2019}, or by using prior knowledge from a CNN pre-trained on a large dataset~\cite{Yue2017,Chen2018,Chen2020} like ImageNet~\cite{Deng2009}. Malware detection was also exploited in the form of a binary classification by considering all malware files as one class and samples of benign software as the other one~\cite{Chen2018,Raff2017}. These networks, however, are vulnerable to adversarial attacks. In this work we explore different different architectures - KNN+GIST as proposed by Nataraj \textit{et. al.} (2011), CNN, CNN-LSTM and CNN BiLSTM as proposed by Le \textit{et. al.} (2018) \cite{Le2018} and MalConv as proposed by Raff (2017) \cite{Raff2017} - and how they behave against crafted adversarial samples.

It is worth mentioning that there are not many relevant public datasets for training malware classifiers, which makes comparing different works a more subtle task. Malimg\cite{Nataraj2011}, BIG 2015\cite{Ronen2018} and EMBER \cite{Anderson2018} are the most notable ones. Since our injection method require reading the file header, BIG 2015\cite{Ronen2018} dataset is not possible because the samples have their headers stripped. EMBER \cite{Anderson2018}, on the other hand, does not provide raw byte values straight away. Since they provide SHA-256 values taken from file contents, a reverse search in malware indexing services is needed in order to retrieve their raw bytes. In Table \ref{tab:related-comparison} we aggregate state of the art methods for malware detection and classification by their technique and the dataset it used.

\begin{table}[ht]
\centering
\renewcommand*\arraystretch{1.3}
\caption{Summary of different malware classification techniques.}
\begin{tabularx}{\linewidth}{|X|X|X|X|}\hline 
  \textbf{AUTHOR} & \textbf{TECHNIQUE} & \textbf{DATASET} \\ \hline
  Nataraj \textit{et. al.} (2011) \cite{Nataraj2011} & GIST + KNN & malimg\cite{Nataraj2011} \\ \hline
  Pascanu \textit{et. al.} (2015) \cite{Pascanu2015} & ESN + Logistic Regresion & Private \\ \hline
  Athiwaratkun e Stokes (2017) \cite{Athiwaratkun2017} & LSTM + MLP & Private \\ \hline
  Yue (2017) \cite{Yue2017} & CNN & malimg\cite{Nataraj2011} \\ \hline
  Raff (2017) \cite{Raff2017} & Embedding + CNN & Private \\ \hline
  Anderson (2018) \cite{Anderson2018} & Embedding + CNN & Ember \cite{Anderson2018} \\ \hline
  Su \textit{et. al.} (2018) \cite{Su2018} & CNN & Private  \\ \hline
  HaddadPajouh \textit{et. al.} (2018) \cite{Haddadpajouh2018} & LSTM & Private \\ \hline
  Liu \textit{et. al.} (2018) \cite{Liu2019} & Multilayer SIFT & malimg\cite{Nataraj2011}, BIG 2015\cite{Ronen2018} \\ \hline
  Agarap e Pepito (2018) \cite{Agarap2018} & GRU + SVM & malimg\cite{Nataraj2011} \\ \hline
  Le (2018) \cite{Le2018} & CNN-BiLSTM & BIG 2015\cite{Ronen2018} \\ \hline
  Chen (2018) \cite{Chen2018} & Inception-V1 \cite{Szegedy2016} & malimg\cite{Nataraj2011}, BIG 2015\cite{Ronen2018} \\ \hline
  Chen (2020) \cite{Chen2020} &  Inception-V1 \cite{Szegedy2016} & Private \\ \hline
  
\end{tabularx}
\label{tab:related-comparison} 
\end{table}


\subsection{Malware injection}
\label{sub:rw-malware-injection}

\begin{figure*}[t]
    \begin{center}
        \begin{tikzpicture}[
            graysquarednode/.style={rectangle, rounded corners, draw=black!80, fill=black!15, very thick, minimum size=5mm},
            bluesquarednode/.style={rectangle, rounded corners, draw=blue!80, fill=blue!15, very thick, minimum size=5mm},
            redsquarednode/.style={rectangle, rounded corners, draw=red!80, fill=red!15, very thick, minimum size=5mm}
        ]

        \node[] (exe) { \includegraphics[width=2cm,height=2cm]{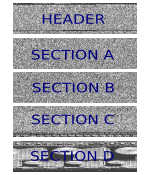} };
        \node[] (header) [below right=0.75cm and 2cm of exe] { \includegraphics[width=2cm,height=2cm]{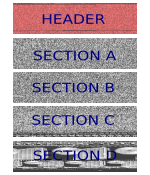} };
        \node[] (ours) [below left=0.75cm and 2cm of exe] { \includegraphics[width=2cm,height=2cm]{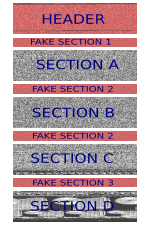} };
        \node[] (padding) [right=0.45cm of ours] { \includegraphics[width=2cm,height=2cm]{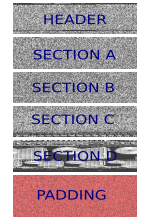} };
        \node[] (modify) [right=0.45cm of padding] { \includegraphics[width=2cm,height=2cm]{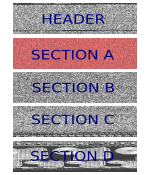} };

        \node[graysquarednode]    (exe_label)    [above=0.0cm of exe]           {\small Input malware};
        \node[bluesquarednode]    (header_label)    [below=0.0cm of header]           {\small \color{blue} Header Manipulation};
        \node[redsquarednode]    (injected_exe_label)    [below=0.0cm of ours]           {\small \color{red} Section Injection};
        \node[bluesquarednode]    (padding_label)    [below=0.0cm of padding]           {\small \color{blue} File Padding};
        \node[bluesquarednode]      (modify_label)        [below=0.0cm of modify]    {\small \color{blue} Content Manipulation};
        
        \draw [-{Triangle[scale=1]}, thick, dashed, draw=blue!80]
            (exe.south)  --  ++(0,-.3) -| (header.north);
        \draw[-{Triangle[scale=1]}, thick, dashed, draw=blue!80]
            (exe.south)  -- ++(0,-.3) -| (padding.north);
        \draw[-{Triangle[scale=1]}, thick, dashed, draw=blue!80]
            (exe.south)  -- ++(0,-.3) -| (modify.north);
        \draw[-{Triangle[scale=1]}, thick, dashed, draw=blue!80]
            (exe.south)  -- ++(0,-.3) -| (ours.north);
        \end{tikzpicture}
        \captionof{figure}{Illustration of the differences among attacks to image-based malware classifiers. Leftmost (red) square displays our approach. Blue squares displays previous attacks.}
        \label{fig:attack-types}
    \end{center}
\end{figure*}
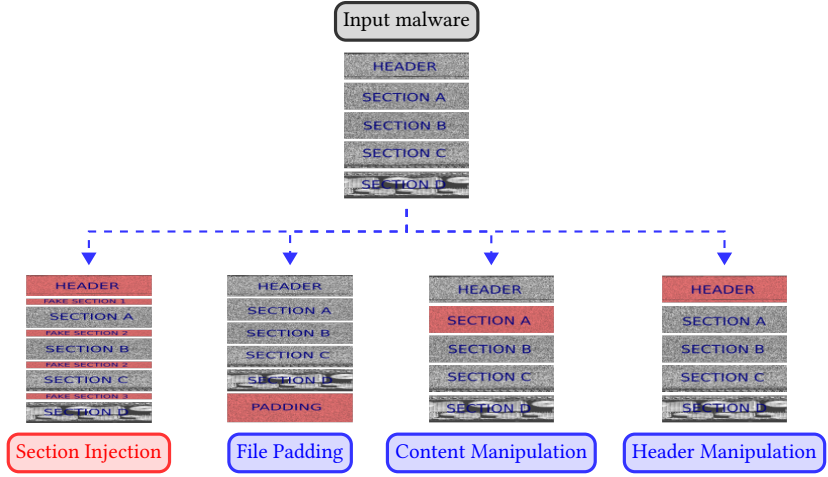

Adversarial attacks consist of adding tiny changes to the input data to alter its classification result and are usually not easily perceived by humans. But, arbitrarily modifying software files without changing its behavior is impossible. Even verifying if a modification does not affect a software's response is an undecidable problem. Thus, if someone arbitrarily alters a malware to change its classification results, there is a chance it will no longer pose a threat to the system. Despite that fact, there exists in literature some possible attacks that retain their functionalities. They are illustrated in Figure~\ref{fig:attack-types}.

Different works exploited adversarial attacks in the malware domain. Grosse~\emph{et~al.}~\cite{Grosse2016} and Al-Dujaili~\emph{et~al.}~\cite{Al-Dujaili2018} extracted static features from malware files and used the Fast Gradient Sign Method (FGSM)~\cite{Goodfellow2015} to modify these feature vectors and form adversarial samples. Notwithstanding, these approaches do not guarantee that it is possible to alter the malware file to produce the adversarial feature vector while maintaining the original functionality. Therefore, they may not have a practical use.

Anderson~\emph{et~al.}~\cite{Anderson2017} explores a black box attack against a reinforcement learning model, where the agent actions are taken from a list of modifications that includes manipulating existing bytes but also adding ones between sections or even creating new sections. No further information is provided regarding the constraints on these injections. It fits "Section Injection" and "Content Manipulation" categories illustrated in Figure~\ref{fig:attack-types}. It achieves evasion rates up to 16\% against a Gradient Boost Decision Tree (\textit{GBDT})~\cite{Anderson2018} model.

Khormali~\emph{et~al.}~\cite{Khormali2019} focused on injecting bytes to the executable files' end, an unreachable area during execution. It fits "File Padding" category illustrated in Figure~\ref{fig:attack-types}. As the operating system will not execute it, not even read it in some cases, it does not affect the malware behavior. These bytes can either be generated by FGSM or be parts of other malware. Nevertheless, extra bytes at the end of the file may be easy to detect and discard before the classification. Besides, this approach requires access to the model or training data used by the classification system, which may not be available in a real attacking scenario.

Demetrio~\emph{et~al.}~\cite{Demetrio2019} proposes a black-box attack called GAMMA (Genetic Adversarial Machine learning Malware Attack), a method that queries a given malware classifier and, based on the output, draws from a set of functionality-preserving manipulations that changes malware samples iteratively. GAMMA is evaluated against two malware classifiers, Malconv\cite{Raff2017} - a shallow neural network - and GBDT\cite{Anderson2018}. Its proposed methods fit all categories illustrated in Figure\ref{fig:attack-types}, despite not detailing how some of those are achieved.

Lucas~\emph{et~al.}~\cite{Lucas2021} also employ functionality preserving techniques. They extend binary rewriting techniques such as in-place randomization (IPR) \cite{Pappas2012} - where the binary is disassembled and some of its instructions are rewritten - and code displacement (Disp) \cite{Koo2016} - where the disassembled version is also used, but with the intent of moving instructions between sections, fitting into "Content Manipulation" category illustrated in Figure~\ref{fig:attack-types}. They apply these attacks in an interactive manner and evaluate them against three neural networks, achieving a misclassification rate of over 80\% in some scenarios.

Benkraouda~\emph{et~al.}~\cite{Benkraouda2021} proposes a framework that mixes a mask generator to highlight the bytes that are possible to manipulate while retaining executability, adversarial example generation using Carlini-Wagner (CW) attack \cite{Carlini2017} and an optimization step that iteratively modifies the masked bytes by comparing the generated adversarial data to a set of known instructions. It fits the "Content Manipulation" category illustrated in Figure\ref{fig:attack-types}. The attack is evaluated against a three-layer CNN, achieving an attack success rate of up to 81.8\%. A shortcoming of this method is the time it takes to generate its samples, reaching over six hours for a single sample in some cases.

Demetrio~\emph{et~al.}~\cite{Demetrio2021} introduce the RAMEN framework, an extensive library with multiple attacks for malware classification. They present three novel attacks - Full DOS, Extend and Shift - all of them capable of modifying the binary sample while keeping its functionality. The novel attacks are evaluated against MalConv~\cite{Raff2017}, DNN with Linear (DNN-Lin) and ReLU (DNN-ReLU)~\cite{Coull2019} and GBDT~\cite{Anderson2018}, being misclassified by the neural networks, but not being able to evade the decision tree since it does not rely only on static data. 

Our attack scheme - Section Injection - is also explored by Anderson~\emph{et~al.}~\cite{Anderson2017} and Demetrio~\emph{et~al.}~\cite{Demetrio2019}, as one possible method in their pipelines, but no further information is provided regarding the constraints for this injection. It can also be seen as an ensemble of \textit{Extend} and \textit{Shift} methods proposed by Demetrio \textit{et al.} \cite{Demetrio2021} and the \textit{padding} methods discussed by Khormali \textit{et al.} \cite{Khormali2019}. The byte modifications presented by Lucas \textit{et al.} \cite{Lucas2021} can also be integrated in our method, leading to the injection of perturbed sections instead of random ones.

Regarding the data used to evaluate the attacks, most works listed here used some sort of private dataset, either by collecting samples from malware hosting services or expanding public ones  - Benkraouda ~\emph{et~al.} \cite{Benkraouda2021} merged malimg\cite{Nataraj2011} and benign samples from Architecture Object Code Dataset (AOCD)\cite{Clemens2015}, Khormali ~\emph{et~al.} \cite{Khormali2019} used BIG 2015\cite{Ronen2018} and also formed a private IoT dataset. A summary of the functionality preserving attacks can be found in Table \ref{tab:injection-methods-comparison}.

\begin{table}[ht]
\centering
\renewcommand*\arraystretch{1.3}
\caption{Summary of functionality preserving attacks against PE32 malware classification.}
\begin{tabularx}{\linewidth}{|X|X|X|X|}\hline 
  \textbf{AUTHOR} & \textbf{METHODS} & \textbf{TARGETS} & \textbf{DATASET} \\ \hline
  Anderson ~\emph{et~al.} \cite{Anderson2017} & Set of Manipulations & \textit{GBDT}\cite{Anderson2018} & Private \\ \hline
  Khormali ~\emph{et~al.} \cite{Khormali2019} & Padding & 3-layer CNN & BIG 2015\cite{Ronen2018} + Private IoT dataset \\ \hline
  Demetrio ~\emph{et~al.} \cite{Demetrio2019} & Set of Manipulations & \textit{MalConv}\cite{Raff2017}, \textit{GBDT}\cite{Anderson2018} & Private \\ \hline
  Demetrio ~\emph{et~al.} \cite{Demetrio2021} & Partial DOS, Full DOS, Extend, Shift, FGSM, Padding & \textit{MalConv}\cite{Raff2017}, \textit{DNN}\cite{Coull2019}, \textit{GBDT}\cite{Anderson2018} & Private \\ \hline
  Lucas ~\emph{et~al.} \cite{Lucas2021} & IPR, Disp & \textit{AvastNet}\cite{Krvcal2018}, \textit{MalConv}\cite{Raff2017}, \textit{GBDT}\cite{Anderson2018} & Private \\ \hline
  Benkraouda ~\emph{et~al.} \cite{Benkraouda2021} & Adversarial Generation + Optimization & CNN\cite{Khormali2019,Kolosnjaji2016} & Private (combination of malimg\cite{Nataraj2011} and AOCD\cite{Clemens2015}) \\ \hline
\end{tabularx}
\label{tab:injection-methods-comparison} 
\end{table}

\section{Data injection}
\label{sec:data-injection}

\begin{figure}[!ht]
    \begin{center}
        \begin{tikzpicture}[
            graysquarednode/.style={rectangle, rounded corners, draw=black!80, fill=black!15, very thick, minimum size=8mm},
            bluesquarednode/.style={rectangle, rounded corners, draw=blue!80, fill=blue!15, very thick, minimum size=8mm},
            redsquarednode/.style={rectangle, rounded corners, draw=red!80, fill=red!15, very thick, minimum size=8mm}
        ]

        \node[] (exe) { \includegraphics[width=1cm,height=1cm]{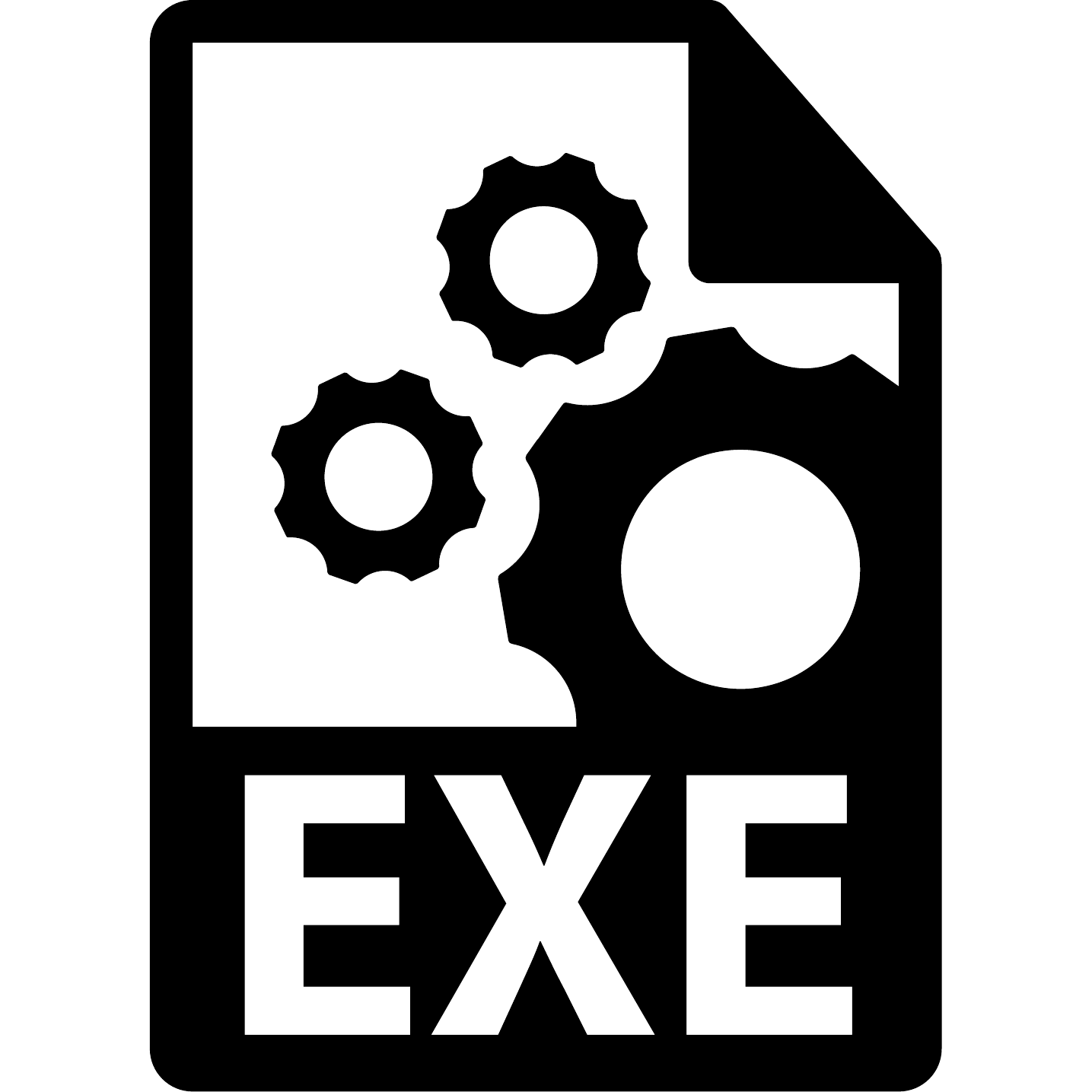}};
        \node[] (img) [right=1.5cm of exe] { \includegraphics[width=1cm,height=1cm]{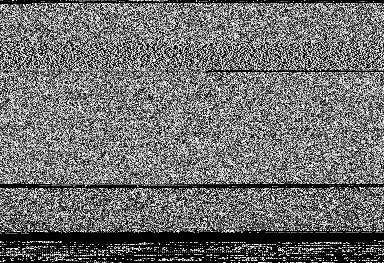} };
        \node[] (injected_exe) [below right=1.5cm and 0.3cm of exe] { \includegraphics[width=1cm,height=1cm]{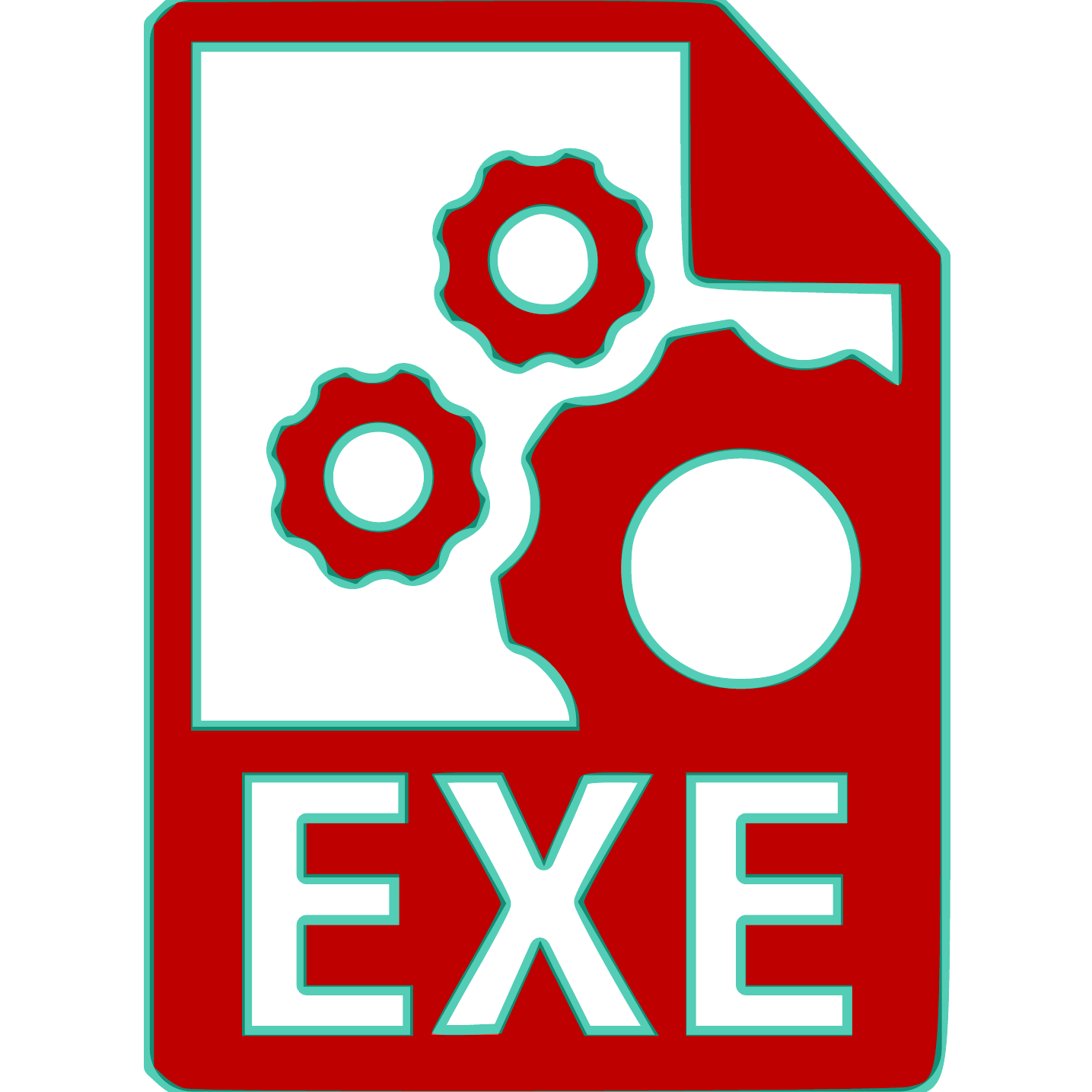} };
        \node[] (classifier) [right=1.2cm of img] { \includegraphics[width=1cm,height=1cm]{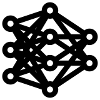} };
        \node[] (family) [below=1.5cm of classifier] { \includegraphics[width=1cm,height=1cm]{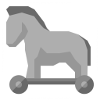} };

        \node[graysquarednode]    (exe_label)    [below=0.0cm of exe]           {\small Input malware};
        \node[bluesquarednode]    (img_label)    [below=0.0cm of img]           {\small \color{blue} Image generation};
        \node[redsquarednode]    (injected_exe_label)    [below=0.0cm of injected_exe]           {\small \color{red} Section injection};
        \node[bluesquarednode]    (classifier_label)    [below=0.0cm of classifier]           {\small \color{blue} Classifier};
        \node[bluesquarednode]      (family_label)        [below=0.0cm of family]    {\small \color{blue} Malware Family};
        
        \draw [-{Triangle[scale=1]}, thick, dashed, draw=blue!80]
            (exe.east)  |-  (img.west);
        \draw[-{Triangle[scale=1]}, thick, draw=red!80, shorten <=5pt]
            (exe_label.south)  |-   (injected_exe.west);
        \draw[-{Triangle[scale=1]}, thick, draw=red!80, shorten >=5pt]
            (injected_exe.east)  -|  (img_label.south);
        \draw[-{Triangle[scale=1]}, thick, draw=blue!80, shorten >=3pt]
            (img.east)  --  (classifier.west);
        \draw[-{Triangle[scale=1]}, thick, draw=blue!80, shorten <=3pt]
            (classifier_label.south)  --  (family.north);
        \draw[-{Triangle[scale=1]}, thick, draw=red!80]
            (injected_exe.east)+(0,-0.25)arc(270:90:-0.8)coordinate(X)
            (injected_exe.west)+(0,-1.85)--(X)
            (injected_exe.west)+(0,-1.85)arc(90:-90:-0.8);
        \end{tikzpicture}
        \captionof{figure}{Flowchart of an image-based malware classification system (blue lines). Red lines replace the dashed blue line in our data injection scheme.}
        \label{system-pipeline}
    \end{center}
\end{figure}
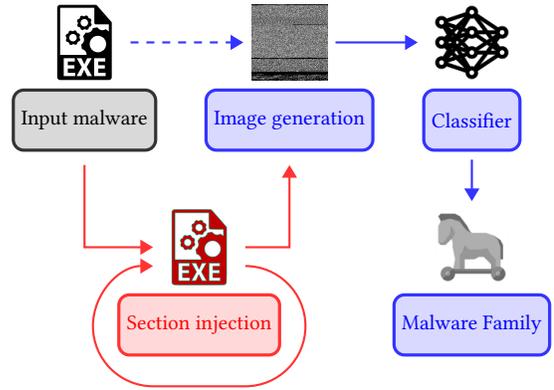

To comply with a realistic usage scenario, we inject one or more sections filled with arbitrary bytes before any processing being done for classification purposes, as illustrated in Figure~\ref{system-pipeline}. We explain how the proposed injection process works in the following sections, and we show how we built the malware classifiers used in our experiments in Section~\ref{sec:malware-classification}.

\subsection{File header}
\label{sub:pe32-header}
The first step in our injection scheme is to obtain information about the input file by reading its header. Table~\ref{tab:pe32-sections} lists the flags that are relevant to us. After inserting a new section, we need to increment the flag $NumberOfSections$ and update the flag $SizeOfImage$ accordingly to preserve the malware functionality. We pick the injected section's index $k$ by drawing a number in the interval $[0,  NumberOfSections]$. Sections $0$ to $k-1$ remain in place, and sections $k$ to $NumberOfSections-1$ are shifted one position forward so that we can insert the new section in $k$-th place.

\begin{table}[ht]
\renewcommand*\arraystretch{1.3}
\centering
\caption{Flags in the header of PE32 files.}
\begin{tabularx}{\linewidth}{|c|X|}\hline
   \textbf{FLAG NAME} & \textbf{DESCRIPTION} \\ \hline
   $NumberOfSections$ & Number of sections in the file \\ \hline
   $FileAlignment$ & Section size in bytes is a multiple of this flag \\ \hline
   $SectionAlignment$ & Memory address of a section is a multiple of this flag \\ \hline
   $SizeOfImage$ & Memory size of all sections in bytes \\ \hline
   $ImageBase$ & Address of the first byte when the file is loaded to memory (default value is 0x00400000) \\ \hline
\end{tabularx}
\label{tab:pe32-sections} 
\end{table}


\subsection{Section header}
\label{sub:pe32-section-header}
A section header is composed of 40 contiguous bytes. These bytes specify what the loader needs to handle this section. Table 2 shows the bytes that we fill when creating a new section. We refer to a flag of the $i$-th section as $Flag_i$.

\begin{table}[ht]
\renewcommand*\arraystretch{1.3}
\centering
\caption{Flags in section headers of PE32 files.}
\begin{tabularx}{\linewidth}{|c|c|X|}\hline
   \textbf{FLAG NAME} & \textbf{SIZE} & \textbf{DESCRIPTION} \\ \hline
   $Name$ & 8 bytes & Section name \\ \hline
   $VirtualSize$ & 4 bytes & Section size in bytes on memory \\ \hline
   $VirtualAddress$ & 4 bytes & Section offset on memory relative to $ImageBase$ \\ \hline
   $SizeOfRawData$ & 4 bytes & Section size in bytes on disk \\ \hline
   $PointerToRawData$ & 4 bytes & Section offset on disk relative to the beginning of the file \\ \hline
   $Characteristics$ & 4 bytes & Section characteristics like usage and permissions \\ \hline
\end{tabularx}
\label{tab:pe32-optional-header} 
\end{table}

First, we generate eight random printable characters (ASCII table values between 33 and 126) as $Name_k$. After that, we set $SizeOfRawData_k$ using Equation~\ref{eq:align-rawdata}:
\begin{equation}
    \small
    \label{eq:align-rawdata}
    SizeOfRawData_k = \lceil\frac{N}{FileAlignment}\rceil \times FileAlignment
\end{equation}
with $N$ being the number of bytes we want to add. We always set $N$ as a multiple of $FileAlignment$ so that null padding is unnecessary. $FileAlignment$ is usually 512 bytes, but it varies according to compilation options.

$PointerToRawData_k$ is set as in Equation \ref{eq:ptrd-last}, if the $k$-th section is the last one. Or else as in Equation \ref{eq:ptrd}, and we add $SizeOfRawData_k$ to $PointerToRawData_i$,
$\forall i > k$.

\begin{equation}
\tiny
\label{eq:ptrd-last}
PointerToRawData_k = PointerToRawData_{k-1} + SizeOfRawData_{k-1}
\end{equation}

\begin{equation}
\tiny
\label{eq:ptrd}
PointerToRawData_k = PointerToRawData_{k+1}
\end{equation}

On memory, we inject sections after every other section to avoid having to update instructions that use memory offsets and preserve the execution path. We set $VirtualAddress_k$ using Equation~\ref{eq:align-virtualaddr}:
\begin{equation}
    \tiny
    \label{eq:align-virtualaddr}
    VirtualAddress_k = \lceil\frac{VirtualAddress_L + VirtualSize_L}{SectionAlignment}\rceil \times SectionAlignment
\end{equation}
where $L$ is the index of the last section on memory. This way, we correctly align the injected section according to $SectionAlingment$.

$VirtualSize_k$ is set to 0, as we do not want to take memory space. Thus, multiple runs of this injection process produce sections pointing to the same address. In our tests, this does not affect execution. We finish our header by setting $Characteristics_k$ as a read-only section with initialized data.

\subsection{Injected data}
\label{sub:pe32-section-payload}
In our work, injected data is a sequence of random bytes. As we have control of the section structure, we could insert pieces from other executables or adversarial examples created using FGSM as other works in the literature~\cite{Khormali2019}. However, we do not do that because we assume we have no access to models and training data used by malware classifiers. Besides, our results show that our simple strategy is enough to affect the performance of a state-of-the-art malware classification approach substantially.

\subsection{Workarounds}
\label{sub:caveats}
We found some challenges when applying this method to an arbitrary PE32 file. Instead of constraining the input files, we dealt with the problems as they appeared. Some malware instances, usually packed or obfuscated, have multiple contiguous virtual sections that do not exist on disk, only on memory. For those cases, we had to adjust the $PointerToRawData$ in injected data to make sure it points to a valid physical section. Furthermore, malware sections are not always correctly aligned with the $FileAlignment$ flag. To avoid fixing existing sections, we only inject data before correctly aligned ones.


\section{Malware classification}
\label{sec:malware-classification}
As can be seen in Figure~\ref{system-pipeline}, this process is divided into two parts: image generation and classification. The former is described in Section~\ref{sub:gen-images}. The latter is carried out with various approaches:

\begin{enumerate}
    \item GIST+KNN \cite{Nataraj2011}, which holds state-of-the-art performance for handcrafted methods
    \item Le-CNN, Le-CNN-LSTM, Le-CNN-BiLSTM \cite{Le2018}, three similar models that uses resizing of the input data to a fixed number of bytes
    \item Malconv \cite{Raff2017}, a model that truncates the first 1MB and perform classification with it.
\end{enumerate}

The trained models are respectively described in Sections~\ref{sub:gist}~and~\ref{sub:cnns}.

\subsection{Image generation}
\label{sub:gen-images}

We transform an executable into an image following Chen's adaptation~\cite{Chen2018} of Nataraj~\emph{et~al.}'s specifications~\cite{Nataraj2011}. We treat every byte as a grayscale pixel, and we break the file into image rows by using a fixed width, which is set according to the file size (see Table~\ref{tab:malware-size}). We discard the last row if it is incomplete. The result is illustrated in Figure~\ref{executable-sections}.

\begin{table}[ht]
\renewcommand*\arraystretch{1.3}
\centering
\caption{Image width based on the executable size \cite{Nataraj2011}.}
\begin{tabular}{|c|c|ccc}
\cline{1-2} \cline{4-5}
{\bf SIZE (kB)} & {\bf WIDTH (px)} & \multicolumn{1}{l|}{} & \multicolumn{1}{c|}{\bf SIZE (kB)} & \multicolumn{1}{c|}{\bf WIDTH (px)} \\ \cline{1-2} \cline{4-5} 
$<$ 10 & 32 & \multicolumn{1}{l|}{} & \multicolumn{1}{c|}{200-500} & \multicolumn{1}{c|}{512} \\ \cline{1-2} \cline{4-5} 
10-30 & 64 & \multicolumn{1}{l|}{} & \multicolumn{1}{c|}{500-1000} & \multicolumn{1}{c|}{768} \\ \cline{1-2} \cline{4-5}
30-60 & 128 & \multicolumn{1}{l|}{} & \multicolumn{1}{c|}{1000-2000} & \multicolumn{1}{c|}{1024} \\ \cline{1-2} \cline{4-5} 
60-100 & 256 & \multicolumn{1}{l|}{} & \multicolumn{1}{c|}{$>$ 2000} & \multicolumn{1}{c|}{2048} \\ \cline{1-2} \cline{4-5} 
100-200 & 384 &                       &                       &                       \\ \cline{1-2}
\end{tabular}
\label{tab:malware-size}
\end{table}


\subsection{Classification}

\subsubsection{GIST+KNN}
\label{sub:gist}
We reproduced Nataraj~\emph{et~al.}'s approach~\cite{Nataraj2011} to the best of our abilities. To do so, we resize our images to $64\times64$ pixels, extract 320-dimensional GIST descriptors, and then classify it using KNN with $K=3$.

\subsubsection{CNNs}
\label{sub:cnns}

Different types of neural networks were explored to classify malware files~\cite{Pascanu2015, Saxe2015,Athiwaratkun2017,Raff2017,Raff2018,Haddadpajouh2018,Su2018,Yue2017,Chen2018, Le2018}. But, to the best of our knowledge, CNNs are the ones with the highest accuracy. In this work we chose different CNN strategies to understand how they perform against data injection: 

\begin{enumerate}
    \item Le \textit{et al}~\cite{Le2018} presents three models. A simple model with three 1D-CNN layers before a fully connected layer, refered to as Le-CNN. A second model with a LSTM layer before the fully connected one, refered to as Le-CNN-LSTM. A third model with a bidirectional LSTM before the fully connected layer, referered to as Le-CNN-BiLSTM. For all of them we employ the same input size of 10k bytes, a batch size of 512, and train the model for at most 60 epochs (early stopping if the accuracy does not improve for 10 epochs).
    \item Raff \textit{et al}~\cite{Raff2017} presents the model refered to as Malconv. This model employs a gated convolution network, i.e., an embedding layer followed by two separate 1D-CNN layers that are multiplied and passed on for two fully connected layers. For this model we use training protocol similar to Lucas \textit{et al} \cite{Demetrio2019}; input size of 1MB, training for a total of 10 epochs without early stopping, with a batch size of 16 due to memory constraints.
\end{enumerate}

\section{Results}
\label{sec:results}

\subsection{Injection Attacks with Random Data}
\label{sub:random-data-attack}

To evaluate malware classification before and after code injection, we use malimg\cite{Nataraj2011} dataset. It has 9,339 malware samples from 25 families. We randomly split the dataset into three parts: training (80\%), validation (10\%), and test (10\%). We use the training and validation sets to perform the CNN training and combine them as a single gallery for the KNN search.

\begin{figure*}
    \centering
   
    \includegraphics[width=0.45\textwidth]{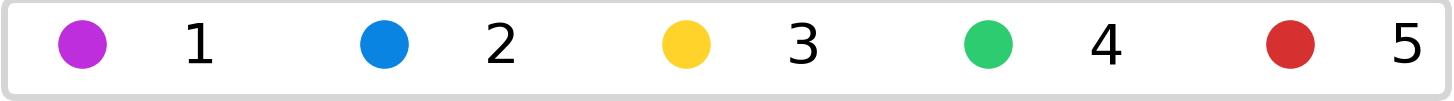}
        
    \begin{subfigure}[t]{0.33\textwidth}
        \centering
        \caption{GIST+KNN}
        \includegraphics[width=\textwidth]{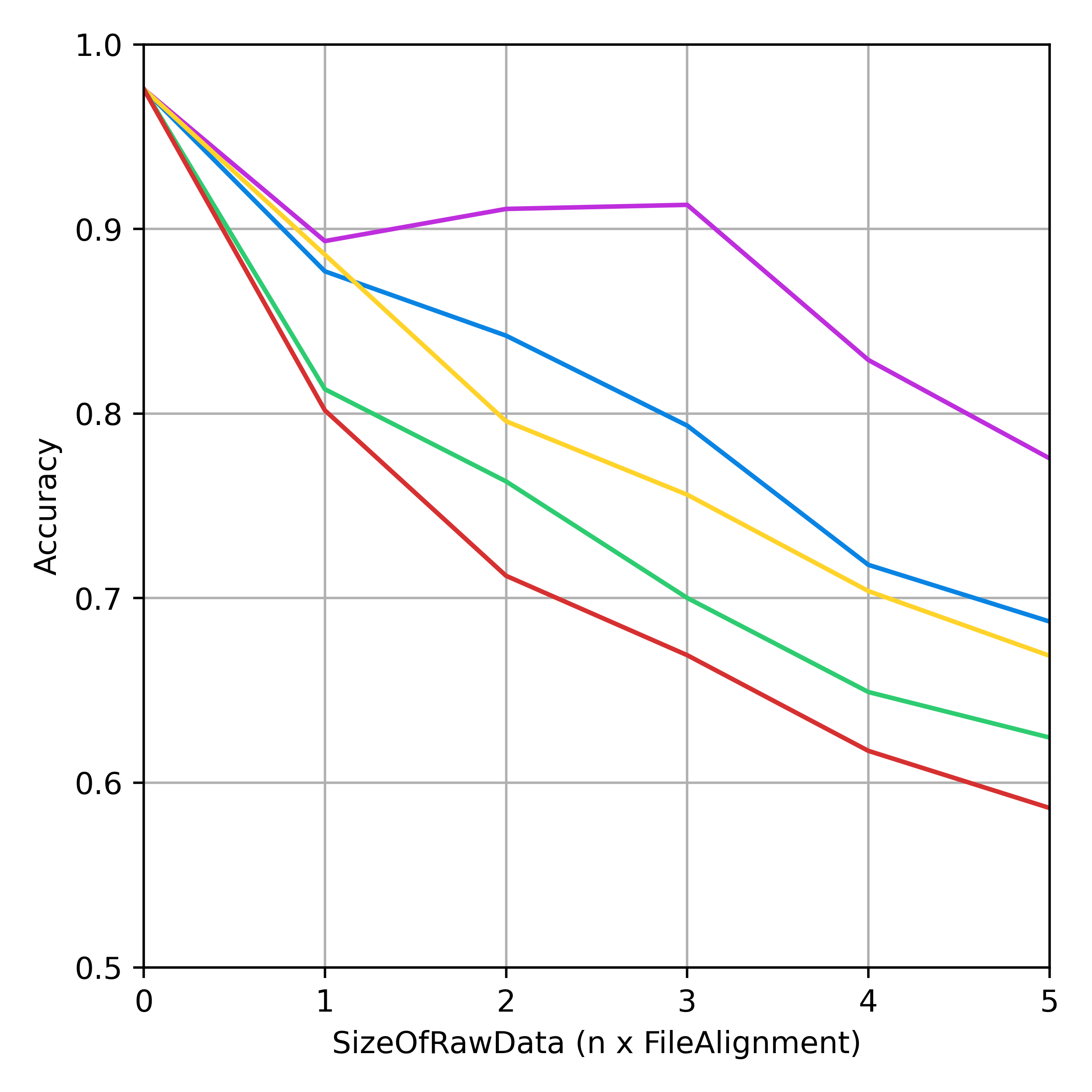}
        \label{acc-plot-gist}
    \end{subfigure}
    \begin{subfigure}[t]{0.33\textwidth}
        \centering
        \caption{MalConv}
        \includegraphics[width=\textwidth]{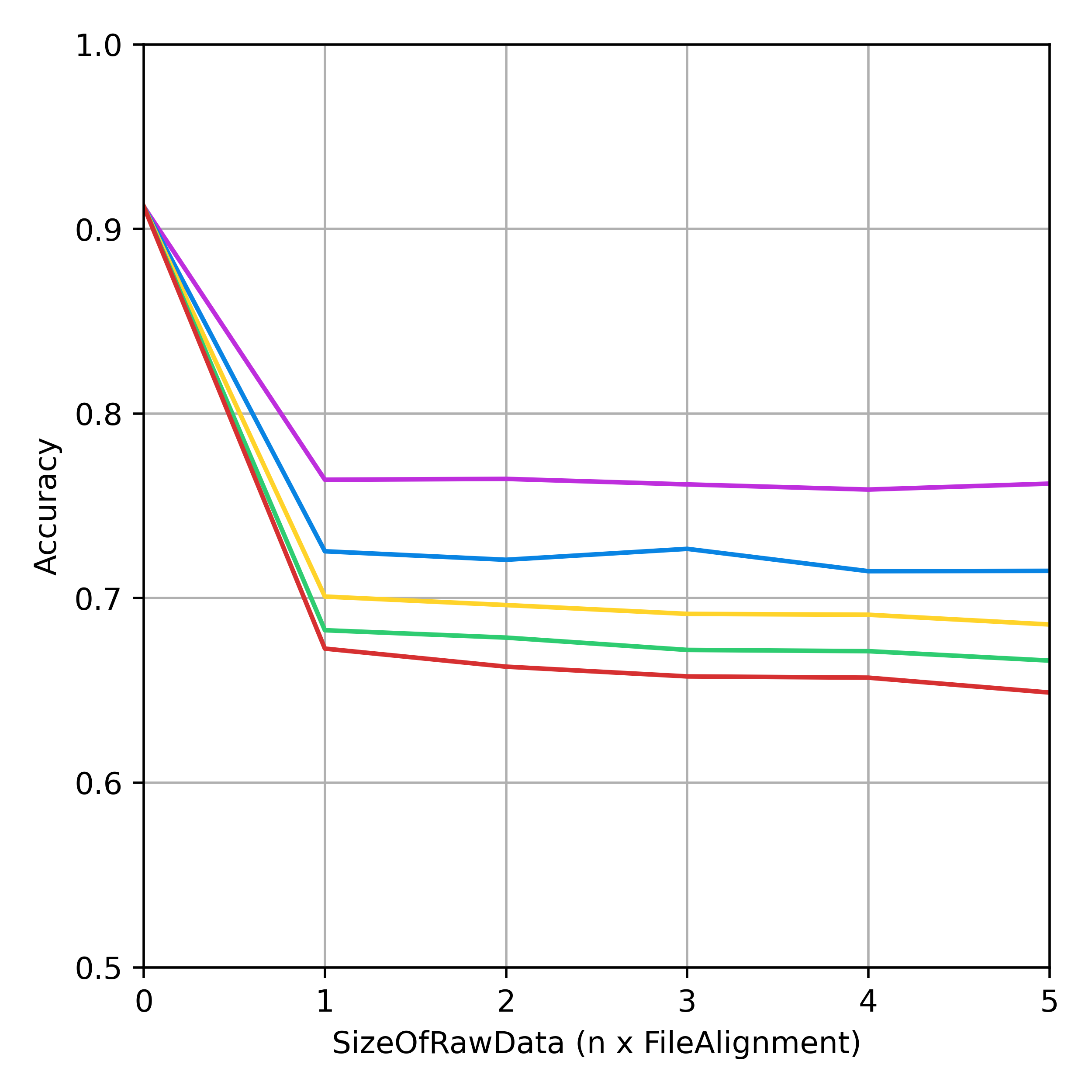}
        \label{acc-plot-malconv}
    \end{subfigure}
    
    \begin{subfigure}[b]{0.33\textwidth}
        \centering
        \caption{Le-CNN}
        \includegraphics[width=\textwidth]{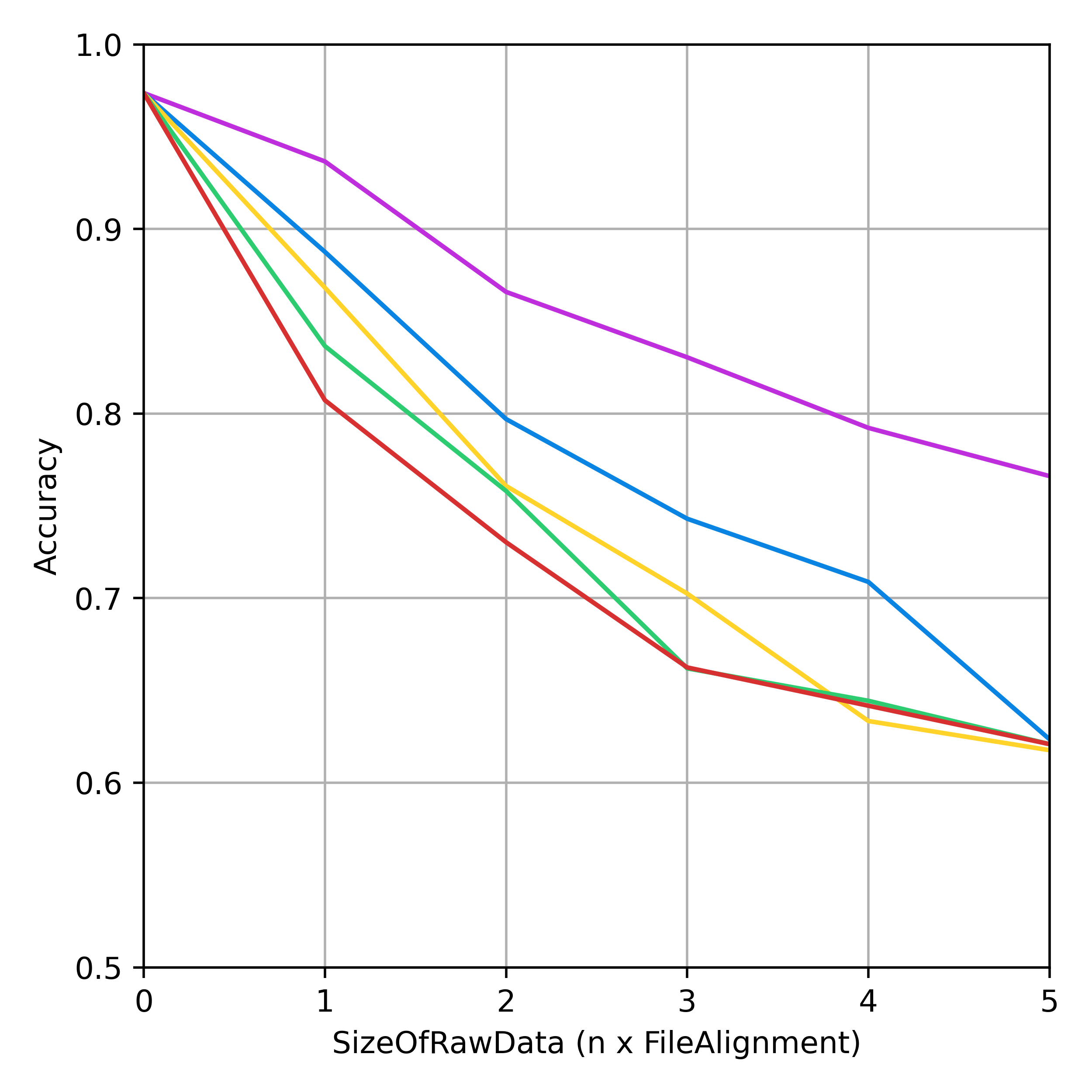}
        \label{acc-plot-cnn}
    \end{subfigure}
    \begin{subfigure}[b]{0.33\textwidth}
        \centering
        \caption{Le-CNN-LSTM}
        \includegraphics[width=\textwidth]{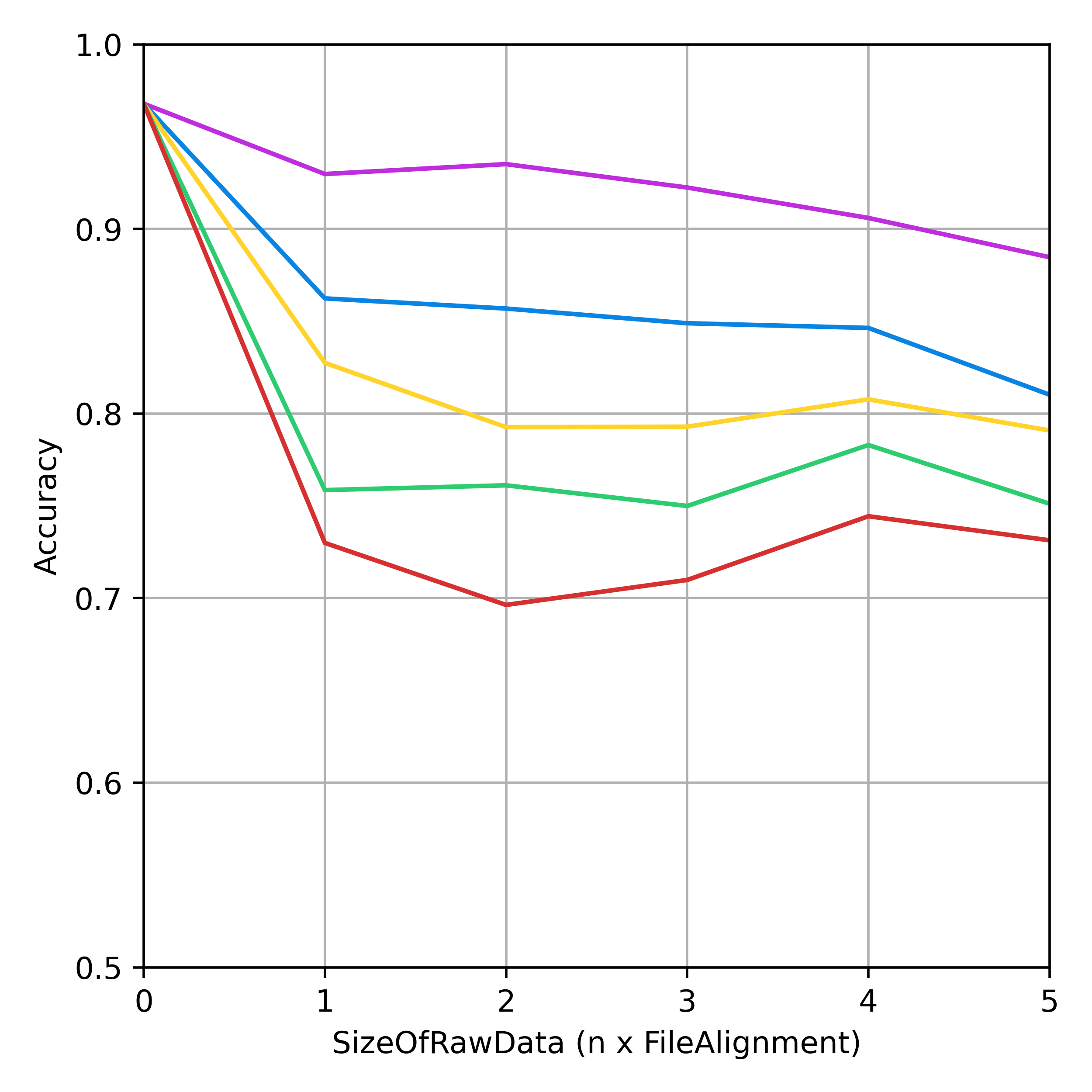}
        \label{acc-plot-cnn-lstm}
    \end{subfigure}
    \begin{subfigure}[b]{0.33\textwidth}
        \centering
        \caption{Le-CNN-BiLSTM}
        \includegraphics[width=\textwidth]{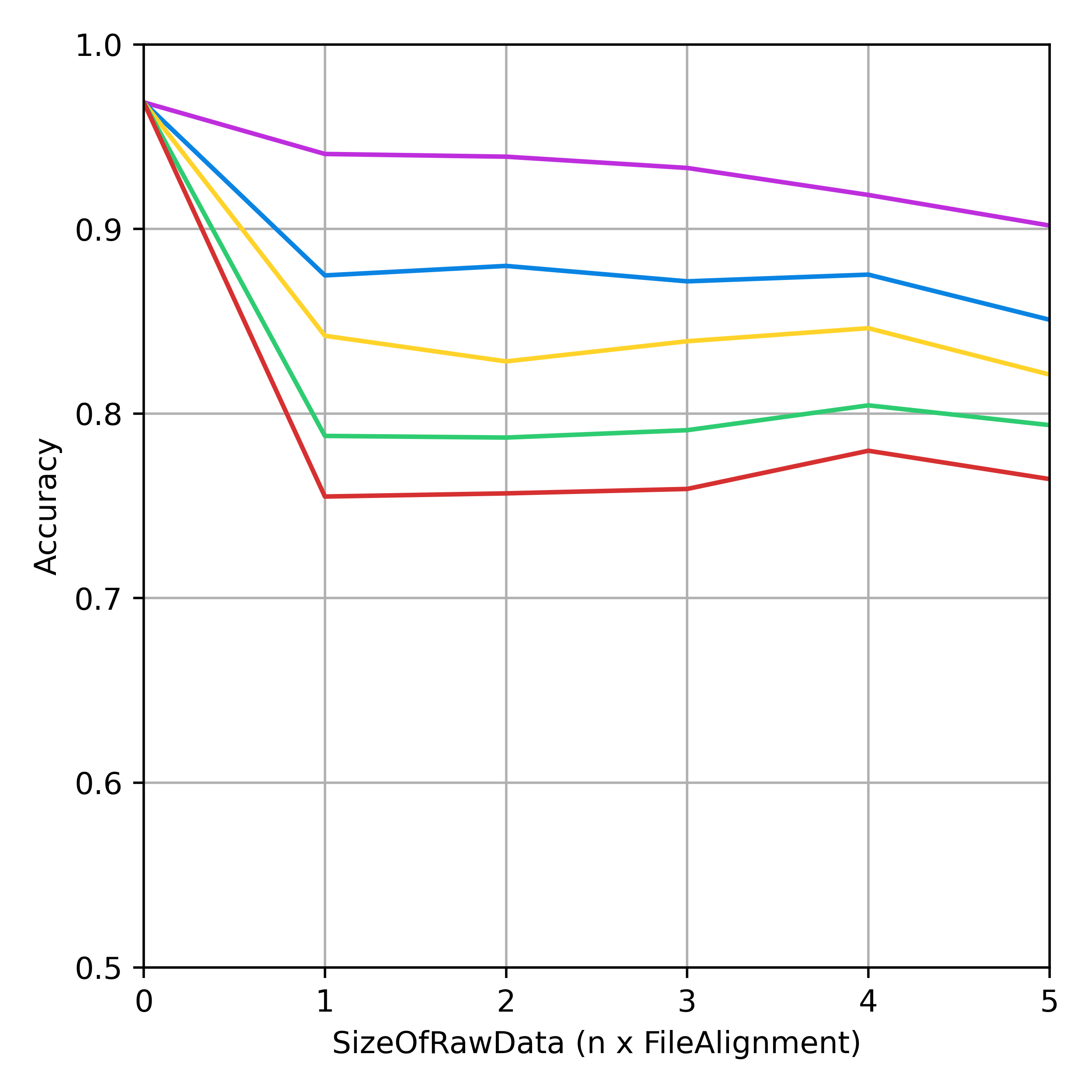}
        \label{acc-plot-cnn-bilstm}
    \end{subfigure}
    
    \caption{Average accuracy of malware classification under different injection scenarios. Different colors represent the amount of injected sections.}
   \label{code-injection-results}
\end{figure*}

For testing, we insert $m$ new sections with $n \times FileAlignment$ bytes at random parts of each test malware, with $m$ and $n$ varying from 1 to 5, totaling 25 different injection scenarios. We repeat training/testing experiments three times for each model and show average results in Figure~\ref{code-injection-results}.

We can see that the way we inject multiple sections affects the results. For instance, despite the amount of data being the same, four sections with $FileAlignment$ bytes impact more the performance than two sections with $2 \times FileAlignment$ or one section with $4 \times FileAlignment$ bytes. Thus, dividing a portion of data in more parts and spreading them over the file is more effective in deceiving the classifier than having a few large sections. 

\begin{figure*}
    \centering
    \begin{subfigure}[b]{0.35\textwidth}
        \centering
        \includegraphics[width=\textwidth]{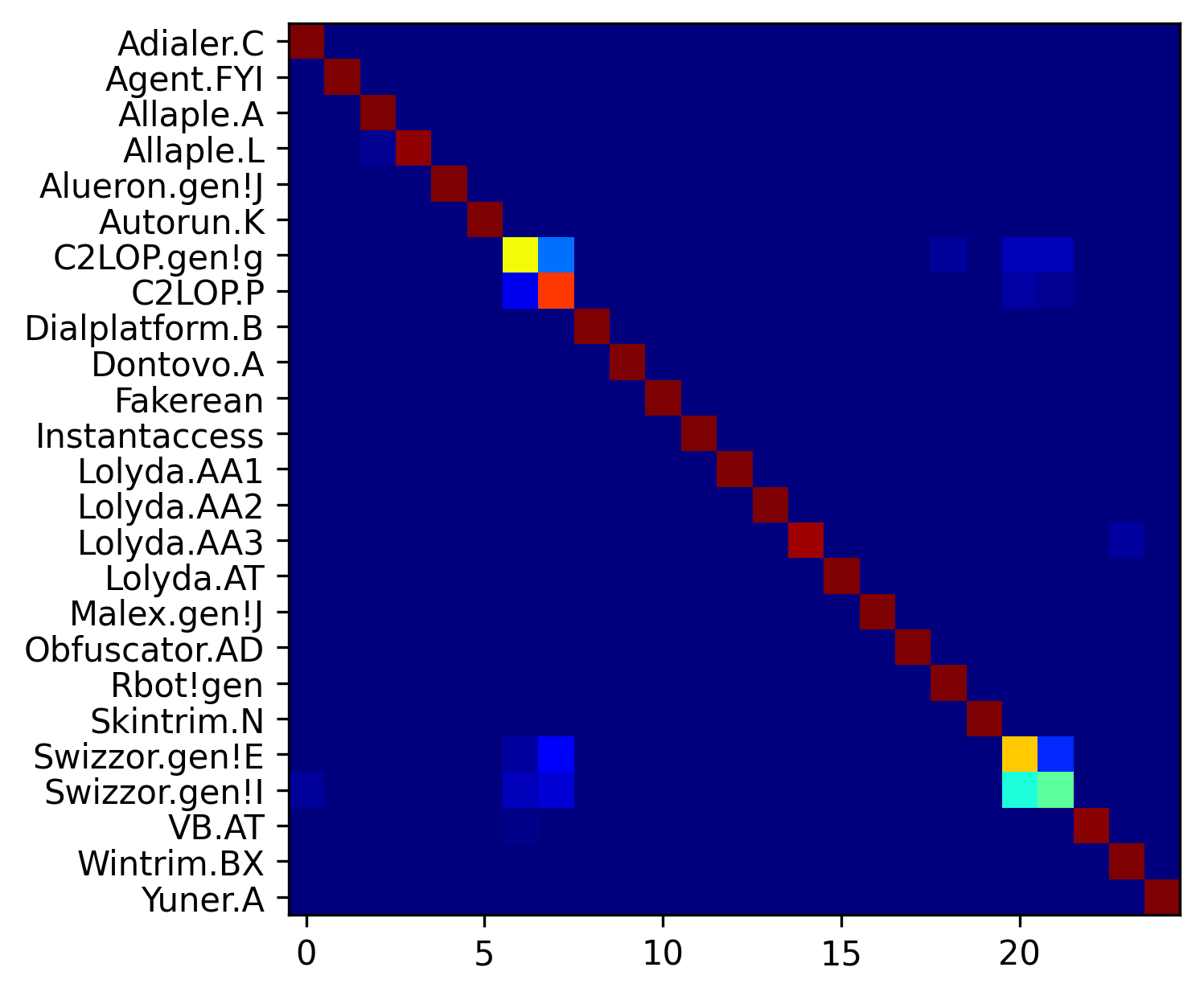}
        \caption{KNN+GIST - Original}
        \label{cm-knn-regular}
    \end{subfigure}
    \begin{subfigure}[b]{0.35\textwidth}
        \centering
        \includegraphics[width=\textwidth,height=0.24\textheight]{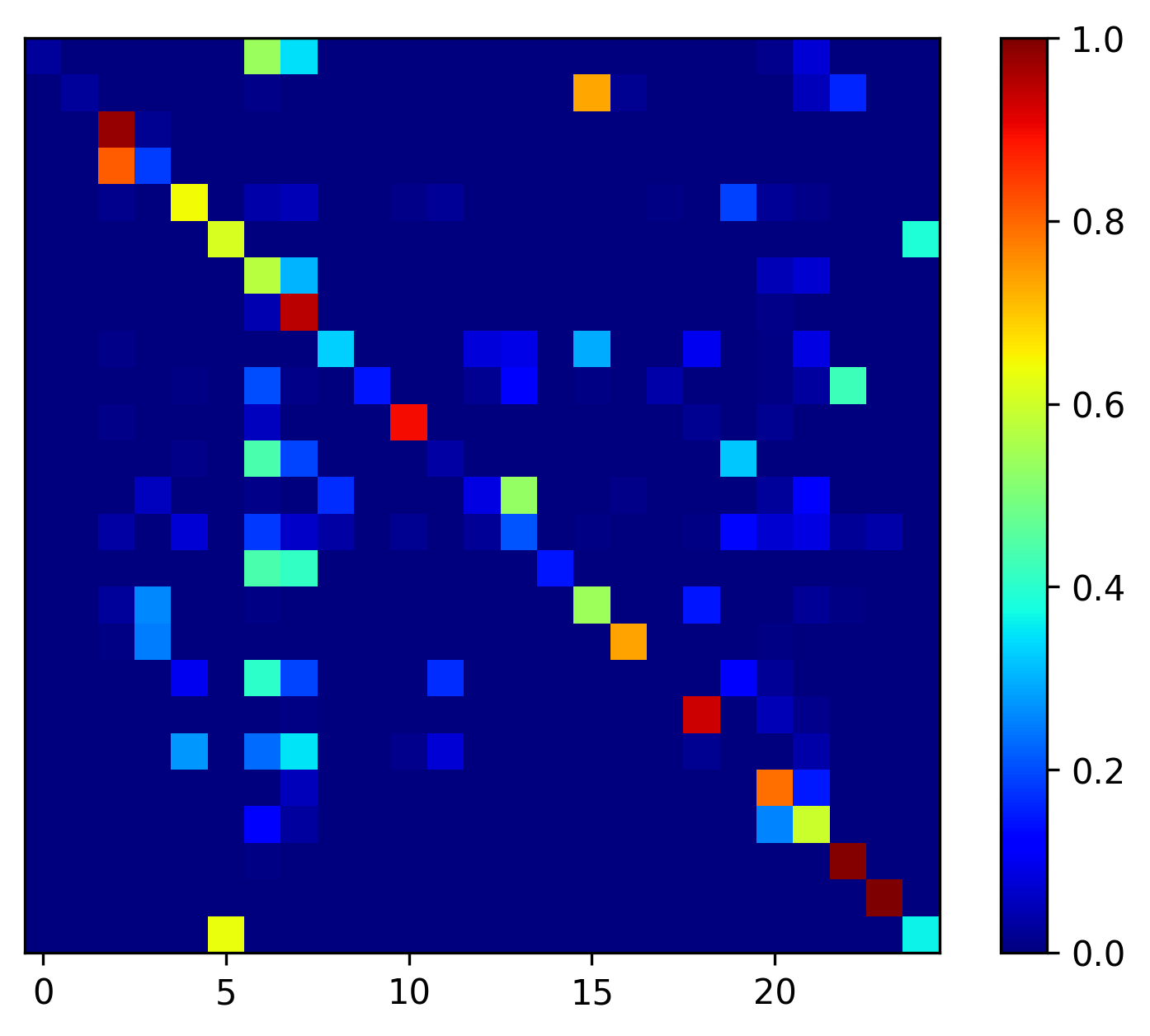}
        \caption{KNN+GIST - Injected}
        \label{cm-knn-injected}
    \end{subfigure}
    
    \begin{subfigure}[b]{0.35\textwidth}
        \centering
        \includegraphics[width=\textwidth]{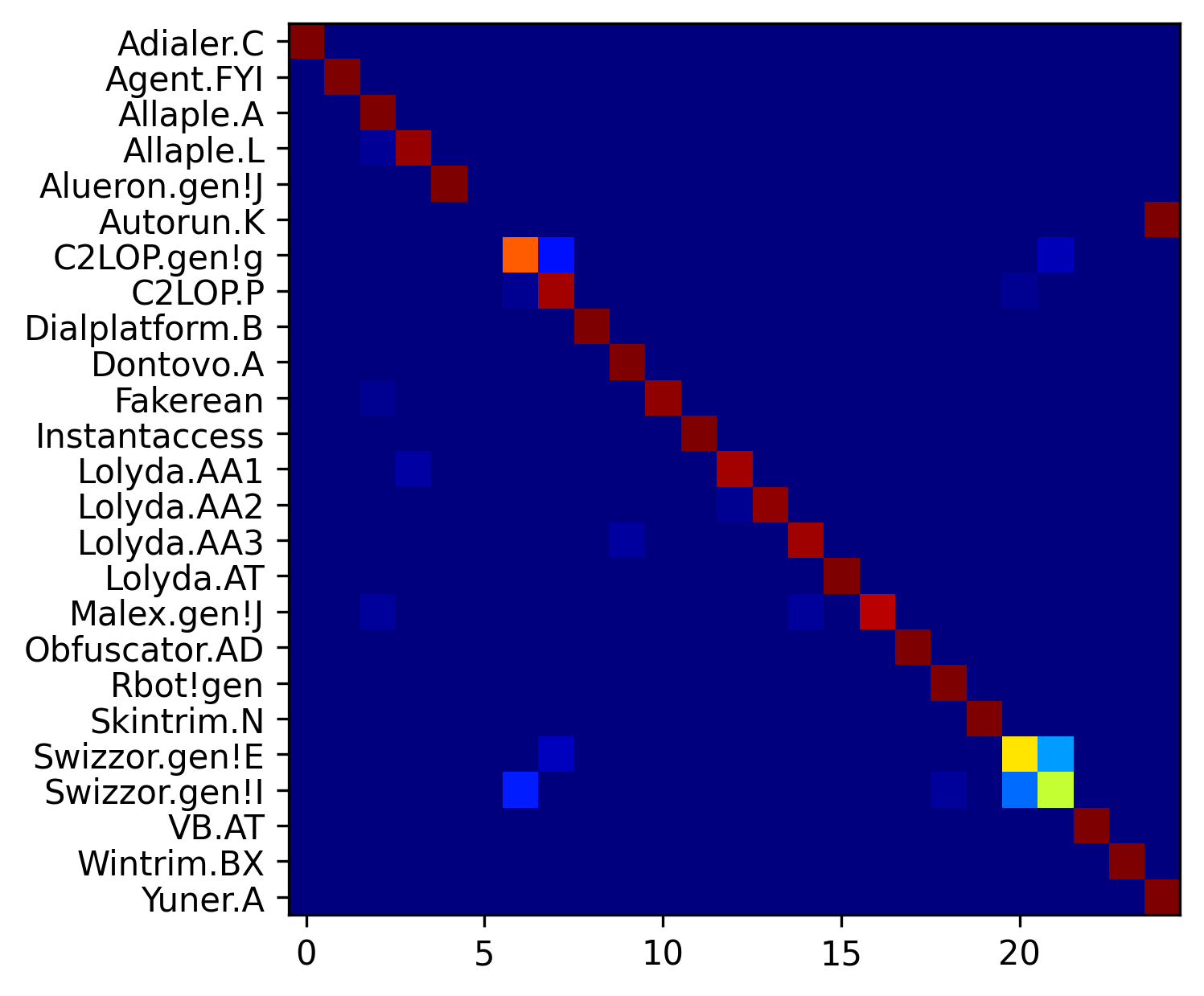}
        \caption{CNN-BiLSTM - Original}
        \label{cm-cnnbilstm-regular}
    \end{subfigure}
    \begin{subfigure}[b]{0.35\textwidth}
        \centering
        \includegraphics[width=\textwidth,height=0.24\textheight]{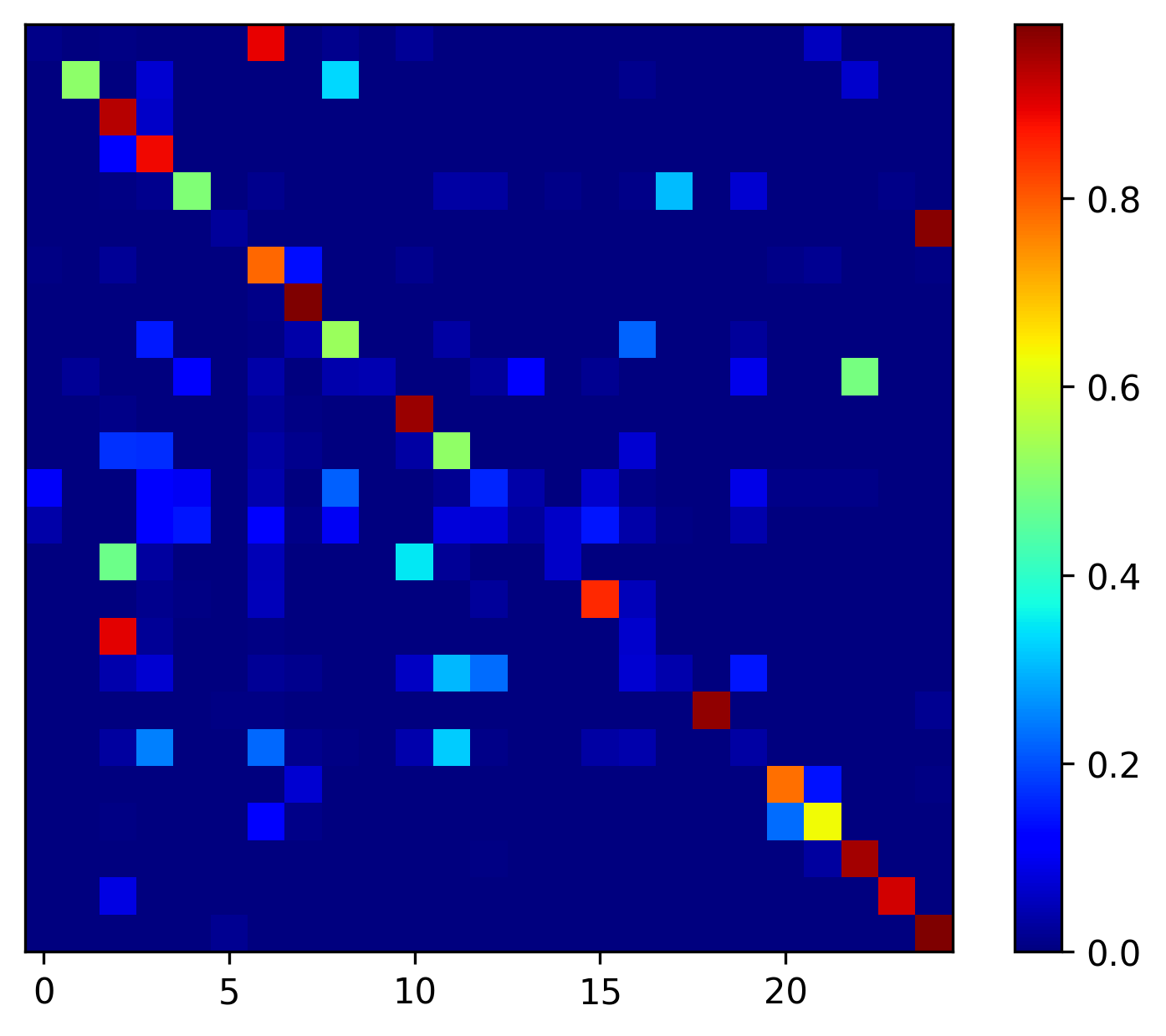}
        \caption{CNN-BiLSTM - Injected}
        \label{cm-cnnbilstm-injected}
    \end{subfigure}
    
    \begin{subfigure}[b]{0.35\textwidth}
        \centering
        \includegraphics[width=\textwidth]{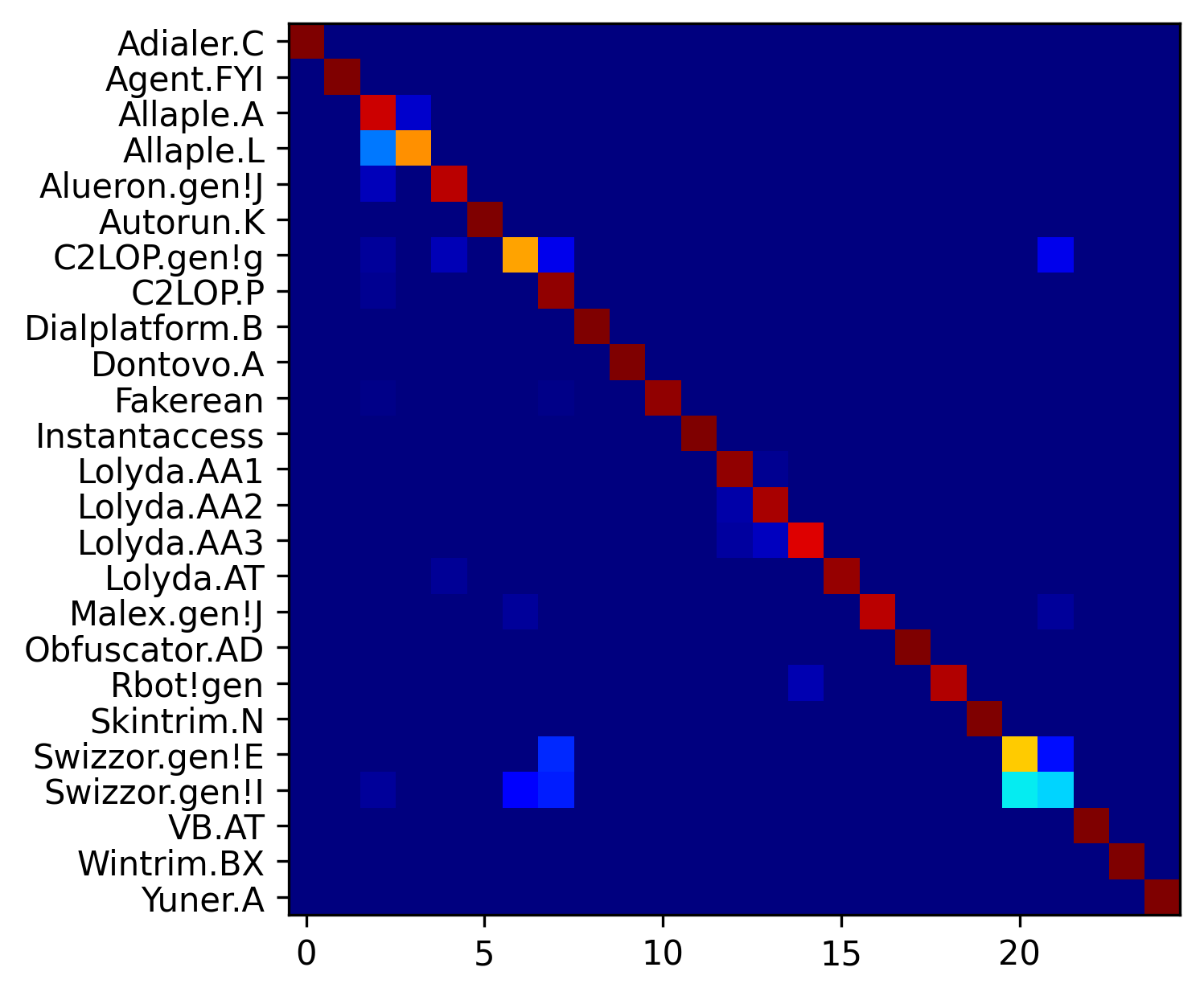}
        \caption{MalConv - Original}
        \label{cm-malconv-regular}
    \end{subfigure}
    \begin{subfigure}[b]{0.35\textwidth}
        \centering
        \includegraphics[width=\textwidth,height=0.24\textheight]{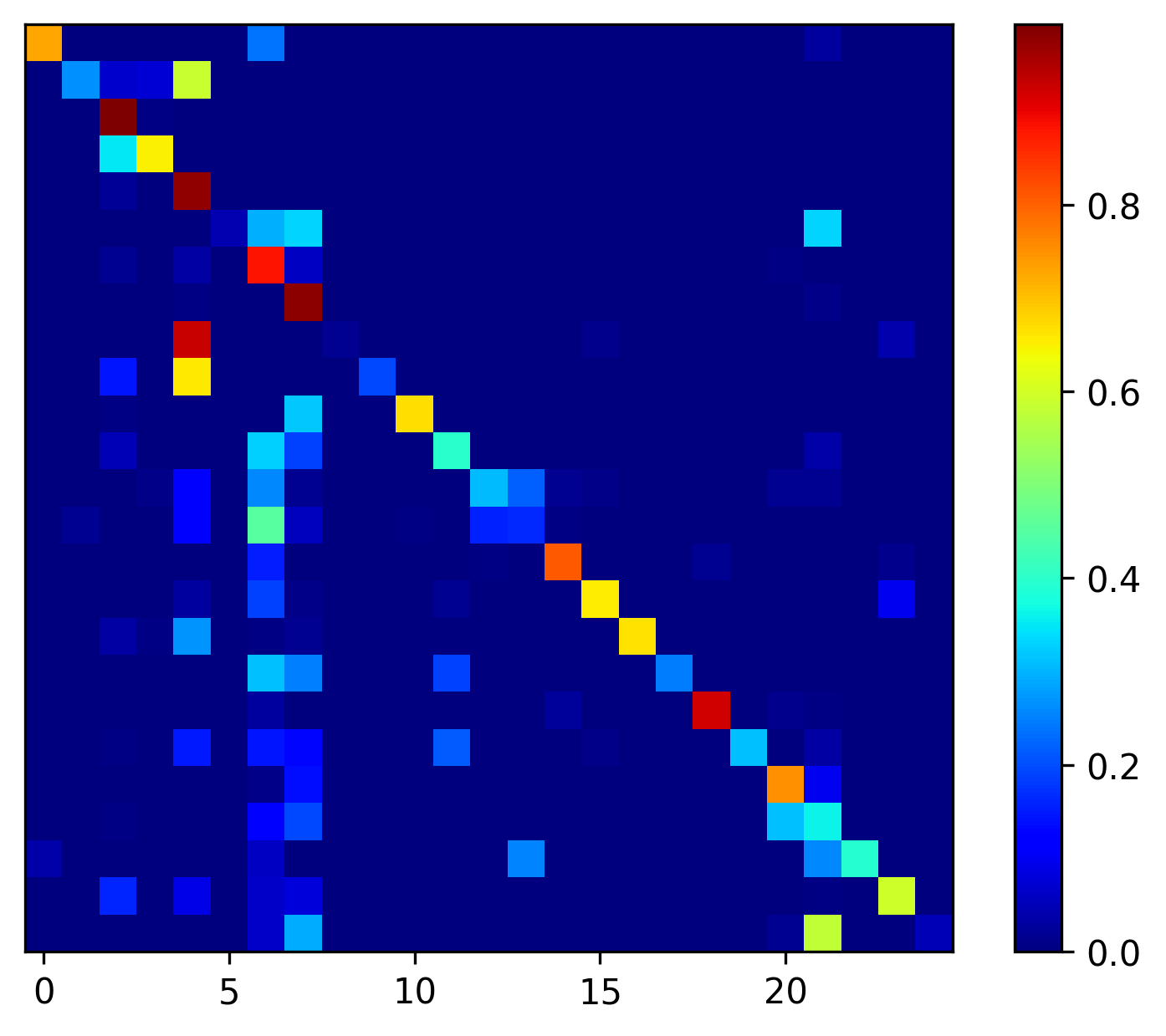}
        \caption{MalConv - Injected}
        \label{cm-malconv-injected}
    \end{subfigure}
    
    \caption{Confusion matrix for malware classification using KNN+GIST, Le-CNN-BiLSTM and MalConv in the original test set~\ref{cm-knn-regular},\ref{cm-cnnbilstm-regular},\ref{cm-malconv-regular} and~\ref{cm-knn-injected},\ref{cm-cnnbilstm-injected},\ref{cm-malconv-injected} when 5 sections of $5 \times FileAlignment$ bytes are injected.}
    \label{confusion-matrix}

\end{figure*}

The biggest drop occurred when we injected five sections with $5 \times FileAlignment$ bytes. As most samples have $FileAlignment = 512$ and the average malware size is 177kB, our injection approach accounts for an approximate 7\% increase in file size and  misclassification rates ranging between 25\% and 40\%. Figure~\ref{confusion-matrix} illustrates the misclassification differences between the test set with original samples and a set with injected samples.


It is worth noting that some of these families share similar traits. For instance, families Autorun.K, Malex.gen!J, Rbot!gen, VB.AT and Yuner.A are all packed using UPX packer. Some families are variants of the same kind of malware, such as C2Lop.P and C2Lop.gen!g, Swizzor.gen!I and Swizzor.gen!E. It is expected that confusion concentrates around those variants \cite{Nataraj2011}.

We can see that all models suffer to correctly classify these variants, even before data injection. Le-CNN-BiLSTM model, as seen in Figures \ref{cm-cnnbilstm-regular} and \ref{cm-cnnbilstm-injected} does not learn how to correctly identify samples from a packed family, \textit{e.g.} "Autorun.K". One behavior is clear in KNN and MalConv models: their tendencies to misclassify samples as "Autorun.K", "C2LOP.gen!g" and "C2LOP.P". Those families share the samples with higher average size in bytes, at 513k, 383k and 523k respectively. Since Le-CNN-BiLSTM resizes everything to 10k bytes, this error is less prevalent with this model. In the same manner, MalConv has these classes as the ones with less misclassifications in the injected set. Considering its 1MB input, those are the samples where padding is used the least.

\begin{figure*}
    \centering
    
    \begin{subfigure}[b]{0.33\textwidth}
        \centering
        \includegraphics[width=\textwidth]{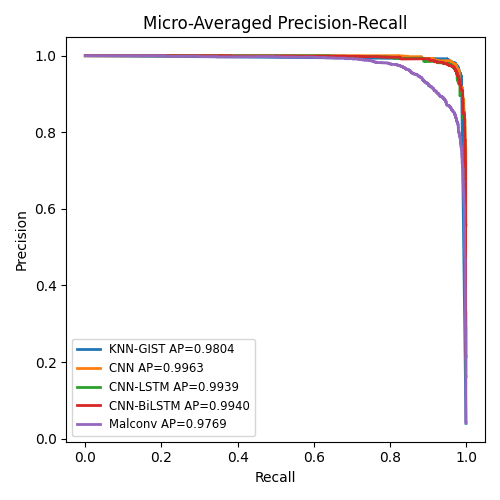}
        \caption{Original}
        \label{pr-plot-regular}
    \end{subfigure}
    \begin{subfigure}[b]{0.33\textwidth}
        \centering
        \includegraphics[width=\textwidth]{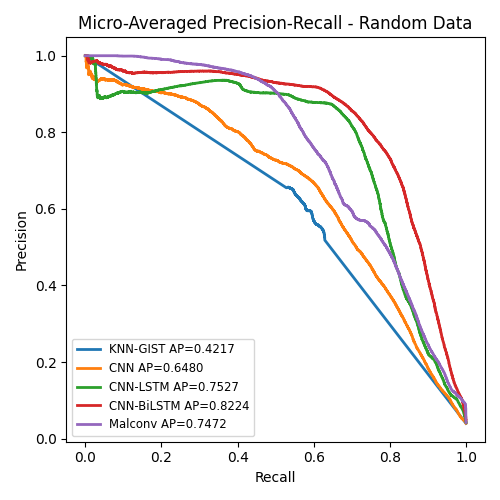}
        \caption{Injected with random data}
        \label{pr-plot-injected}
    \end{subfigure}
    \begin{subfigure}[b]{0.33\textwidth}
        \centering
        \includegraphics[width=\textwidth]{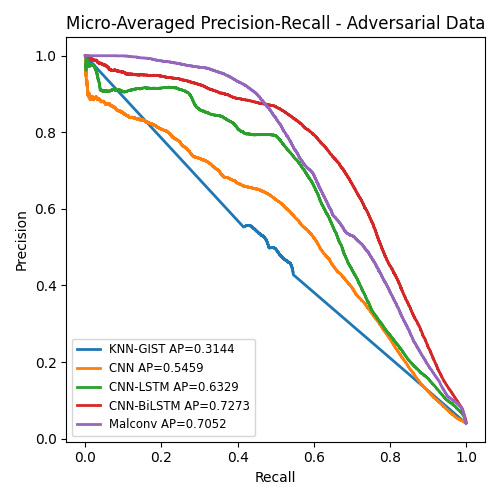}
        \caption{Injected with adversarial data}
        \label{pr-adversarial-injected}
    \end{subfigure}
    
    \caption{Precision Recall curves in the original test set~\ref{pr-plot-regular} and when 5 sections of $5 \times FileAlignment$ bytes are injected, \ref{pr-plot-injected} with random bytes and adversarial bytes in \ref{pr-adversarial-injected}. Each color represents a different model.}
    \label{pr-injection-results}

\end{figure*}

In Figure \ref{pr-injection-results} we can see how the trained models lose precision after section injection. Due to the imbalanced nature of the dataset, this is illustrated by Precision-Recall curves.

We can see that the handcrafted method is the least precise in this scenario, being followed by Le-CNN, MalConv, Le-CNN-LSTM and Le-CNN-BiLSTM respectively.

\subsection{Injection Attacks with Adversarial Data}
\label{sub:adversarial-data-attack}

What if instead of adding random data we use bytes that appear in samples from other classes? We evaluate this kind of attack in this section, this time focusing on the most impactful injection scenario, i.e., 5 sections with $5 \times FileAlignment$. Figure \ref{pr-adversarial-injected} displays the difference that injecting with adversarial data imposes.

Comparing with random injection results seen in Figure \ref{pr-plot-injected}, we can see that all models had its average precision decreased - KNN+GIST by 25.44\%, Le-CNN by 15.75\%, Le-CNN-LSTM by 15.91\%, Le-CNN-BiLSTM by 11.56\% and MalConv by 5.62\%. That might be an indication that MalConv is learning more discriminative features from the samples and it is deceived for reasons other than the kind of data being injected, since it becomes the model with highest average precision despite losing more accuracy than Le-CNN-BiLSTM (Figure \ref{code-injection-results}).

\subsection{Evaluating Samples without Header}
\label{sub:headerless-data-attack}

\begin{figure*}
    \centering
    
    \begin{subfigure}[b]{0.35\textwidth}
        \centering
        \includegraphics[width=\textwidth]{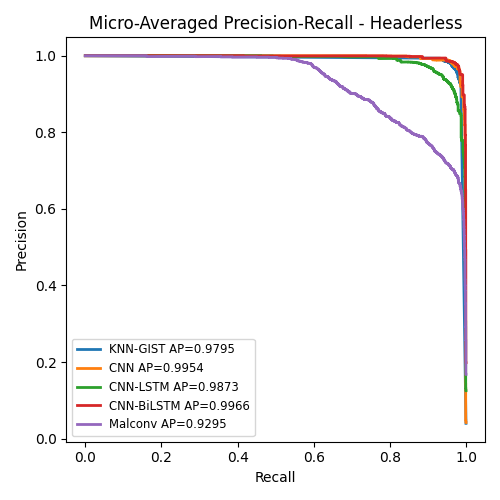}
        \caption{Original}
        \label{pr-plot-headerless-regular}
    \end{subfigure}
    \begin{subfigure}[b]{0.35\textwidth}
        \centering
        \includegraphics[width=\textwidth]{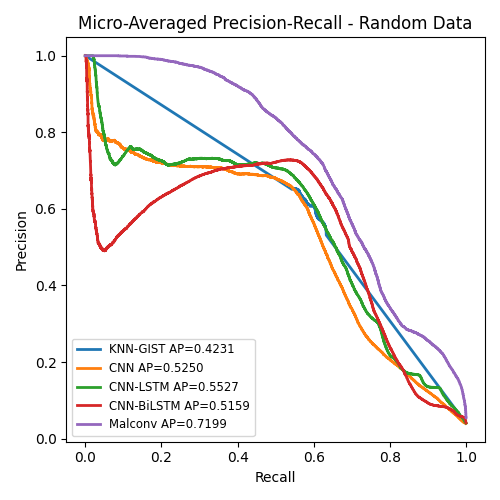}
        \caption{Injected}
        \label{pr-plot-headerless-injected}
    \end{subfigure}
    
    \caption{Precision Recall curves in the original test set~\ref{pr-plot-headerless-regular} and~\ref{pr-plot-headerless-injected} when 5 sections of $5 \times FileAlignment$ bytes are injected, both versions without file header.}
    \label{headerless-injection-results}
    
\end{figure*}

Here we evaluate the possibility of training our models striping the header of the samples, similarly to what is employed in BIG 2015\cite{Ronen2018}. Figure \ref{headerless-injection-results} illustrates the results for samples without the header. 

All models rely heavily on the samples header in order to perform classification, losing precision even before data injection as seen in \ref{pr-plot-headerless-regular}. Only Le-CNN-BiLSTM increased its precision by 0.0026 in this scenario. All models became less robust to data injection, losing precision significantly when compared to complete executables in \ref{pr-plot-injected}. Despite that, MalConv is the only model with similar average precision to previous scenarios.


\subsection{Defending Against Data Injection}
\label{sub:results-defense}

Multiple strategies were evaluated in order to make these models more robust against data injection, by focusing on the kind of data available during training.

\subsubsection{Augmentation}

\begin{figure*}
    \centering
    
    \begin{subfigure}[b]{0.33\textwidth}
        \centering
        \includegraphics[width=\textwidth]{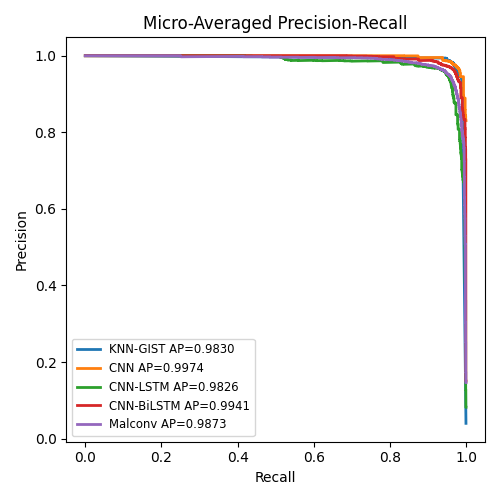}
        \caption{Reordering x Unmodified}
        \label{pr-defense-reordering}
    \end{subfigure}
    \begin{subfigure}[b]{0.33\textwidth}
        \centering
        \includegraphics[width=\textwidth]{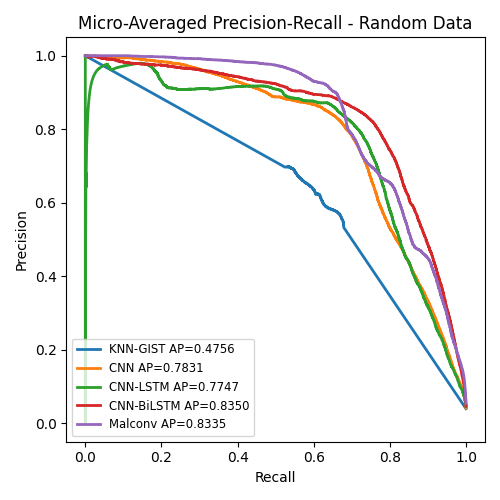}
        \caption{Reordering x Random}
        \label{pr-defense-reordering-random}
    \end{subfigure}
    \begin{subfigure}[b]{0.33\textwidth}
        \centering
        \includegraphics[width=\textwidth]{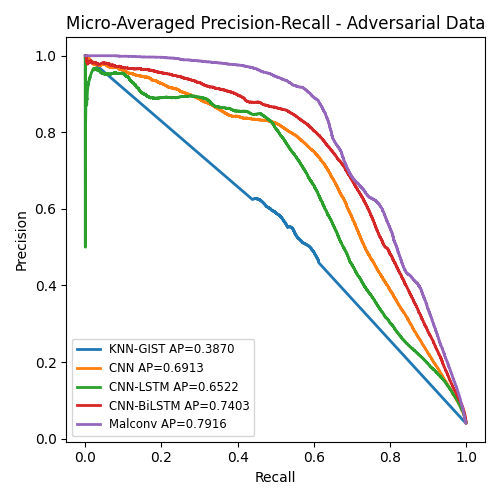}
        \caption{Reordering x Adversarial}
        \label{pr-defense-reordering-adv}
    \end{subfigure}
    
    \begin{subfigure}[b]{0.33\textwidth}
        \centering
        \includegraphics[width=\textwidth]{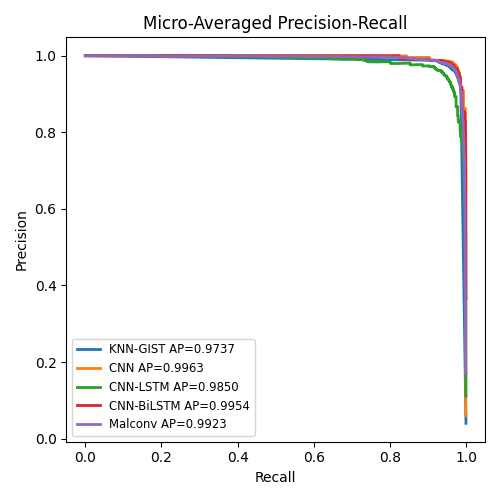}
        \caption{Injection x Unmodified}
        \label{pr-defense-injection}
    \end{subfigure}
    \begin{subfigure}[b]{0.33\textwidth}
        \centering
        \includegraphics[width=\textwidth]{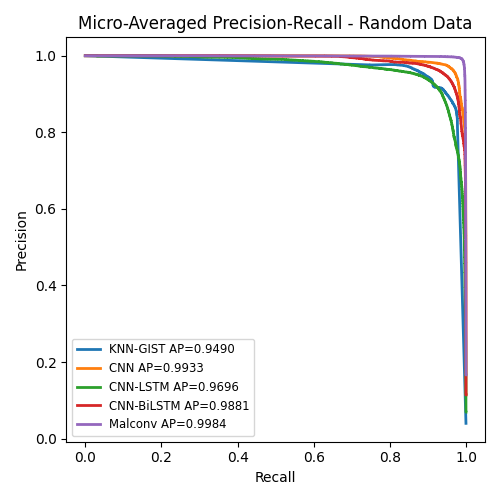}
        \caption{Injection x Random}
        \label{pr-defense-injection-random}
    \end{subfigure}
    \begin{subfigure}[b]{0.33\textwidth}
        \centering
        \includegraphics[width=\textwidth]{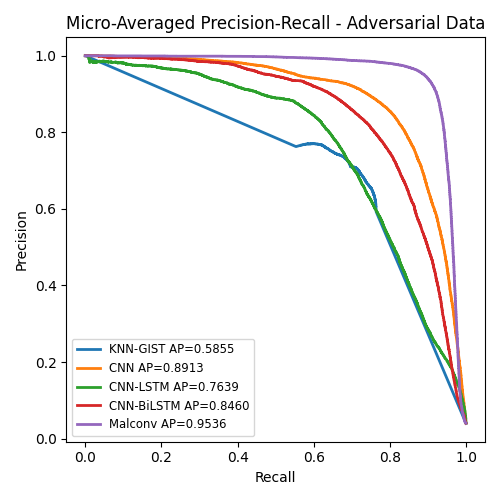}
        \caption{Injection x Adversarial}
        \label{pr-defense-injection-adv}
    \end{subfigure}
        
    \begin{subfigure}[b]{0.33\textwidth}
        \centering
        \includegraphics[width=\textwidth]{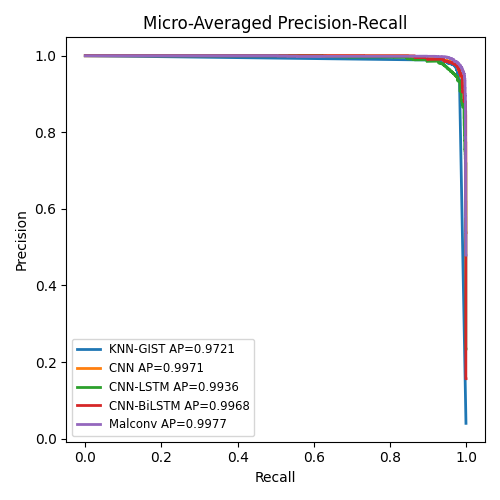}
        \caption{Reordering + Injection x Unmodified}
        \label{pr-defense-augmentation}
    \end{subfigure}
    \begin{subfigure}[b]{0.33\textwidth}
        \centering
        \includegraphics[width=\textwidth]{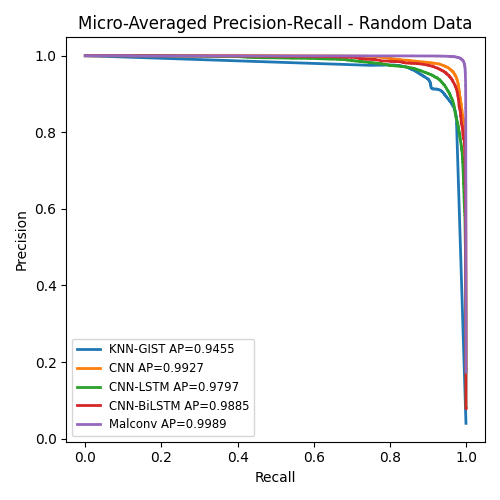}
        \caption{Reordering + Injection x Random}
        \label{pr-defense-augmentation-random}
    \end{subfigure}
    \begin{subfigure}[b]{0.33\textwidth}
        \centering
        \includegraphics[width=\textwidth]{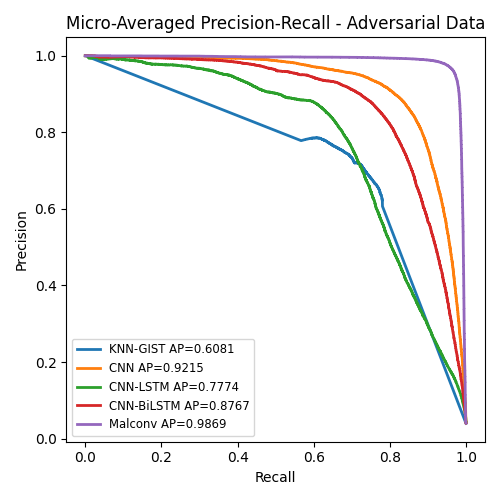}
        \caption{Reordering + Injection x Adversarial}
        \label{pr-defense-augmentation-adv}
    \end{subfigure}
    
    \caption{Precision Recall curves for tests with augmentation. \ref{pr-defense-reordering}, \ref{pr-defense-injection} and \ref{pr-defense-augmentation} displays results on test sets with original samples. \ref{pr-defense-reordering-random}, \ref{pr-defense-injection-random} and \ref{pr-defense-augmentation-random} displays results for datasets injected with random data. \ref{pr-defense-reordering-adv}, \ref{pr-defense-injection-adv} and \ref{pr-defense-augmentation-adv} display results for injection with adversarial data. 5 sections of $5 \times FileAlignment$ bytes are injected in all cases. Each color represents a different model.}
    \label{pr-defense-results}

\end{figure*}

A solution proposed in the literature \cite{Catak2021, Perez2017, Taylor2018} is to augment the data used for training. Three strategies were initially evaluated:

\begin{enumerate}
    \item \textbf{Section reordering:} since our injection scheme adds new sections in a random position among the existing one, the first augmentation idea was to reorder the sections on the training section. By doing this we wanted to check if the model could be more robust against data injection without seeing them during training. As shown by Figures \ref{pr-defense-reordering}, \ref{pr-defense-reordering-random} and \ref{pr-defense-reordering-adv}, this strategy increased a bit the performance of all models when compared to the vanilla results shown by Figure \ref{pr-injection-results}.
    \item \textbf{Training with injected data:} since data is injected in the test set, a possibility was to include injected samples with random data in the training set as well. By comparing Figures \ref{pr-defense-injection}, \ref{pr-defense-injection-random}, \ref{pr-defense-injection-adv} against Figure \ref{pr-injection-results} we can see that all models became less vulnerable against random data injection, but still struggle against adversarial data. MalConv was benefitted the most in this scheme.
    \item \textbf{Reorder+Injection:} augmentating the training set with both injected and reordered samples, shown in Figures \ref{pr-defense-augmentation}, \ref{pr-defense-augmentation-random}, \ref{pr-defense-augmentation-adv}, was also evaluated. Comparing with the original results in Figure \ref{pr-injection-results} we can see that this may be a good defense strategy as well.
\end{enumerate}

As shown by Figure \ref{pr-defense-results}, some models were improved by these augmentation strategies, even tough still vulnerable.

\subsubsection{Binary Data}

\begin{figure*}
    \centering
 
    \begin{subfigure}[b]{0.33\textwidth}
        \centering
        \includegraphics[width=\textwidth]{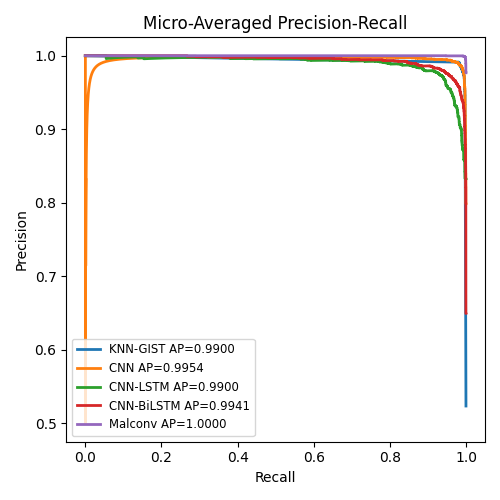}
        \caption{Results on unmodified binary dataset}
        \label{pr-defense-binary-regular}
    \end{subfigure}
    \begin{subfigure}[b]{0.33\textwidth}
        \centering
        \includegraphics[width=\textwidth]{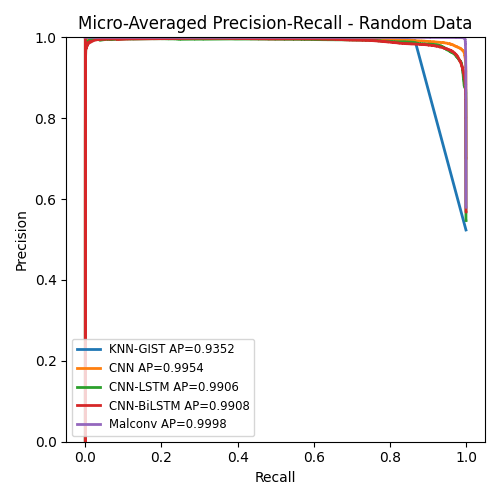}
        \caption{Results on random injected malware}
        \label{pr-defense-binary-random}
    \end{subfigure}
    \begin{subfigure}[b]{0.33\textwidth}
        \centering
        \includegraphics[width=\textwidth]{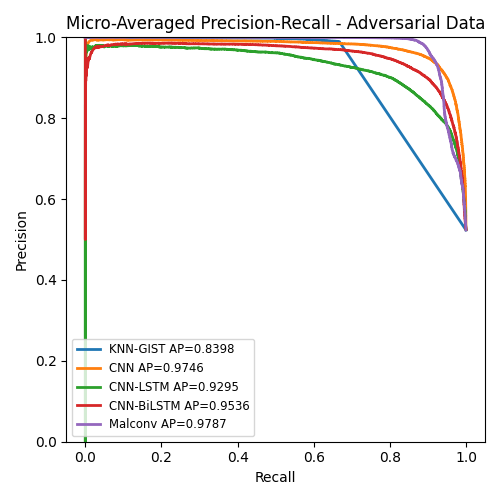}
        \caption{Results on adversarial-injected malware}
        \label{pr-defense-binary-adv}
    \end{subfigure}
    
    \caption{Precision Recall curves when 5 sections of $5 \times FileAlignment$ bytes are injected, \ref{pr-defense-binary-random} with random bytes and adversarial bytes in \ref{pr-defense-binary-adv}. Each color represents a different model.}
    \label{pr-defense-binary-results}

\end{figure*}

All experiments mentioned here were also performed in a binary dataset. We collected samples from a clean Windows 10 Virtual Machine. In this dataset the models were barely affected by data injection. We believe something similar to the mentioned by Raff \textit{et al} \cite{Raff2017} also happened in our dataset: models were learning "Microsoft vs Non-Microsoft" instead of "Benign vs Malign", as shown by Figure \ref{pr-defense-binary-results}. For these tests we evaluated the models performance on the original test set and against malware-only versions of the dataset injected with both random and adversarial data.

\subsubsection{Scaling models}

\begin{figure}[!ht]
    \begin{center}
        \includegraphics[width=\linewidth]{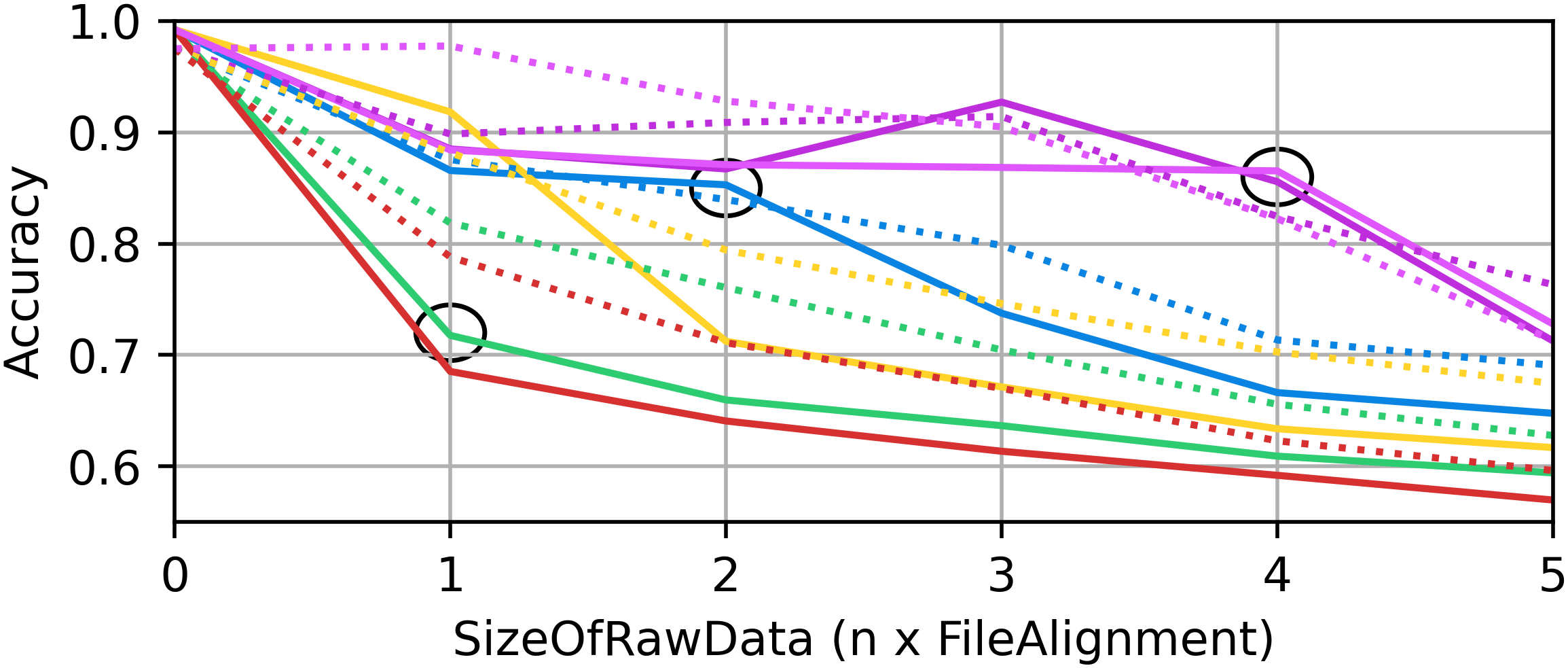}
        \includegraphics[width=\linewidth]{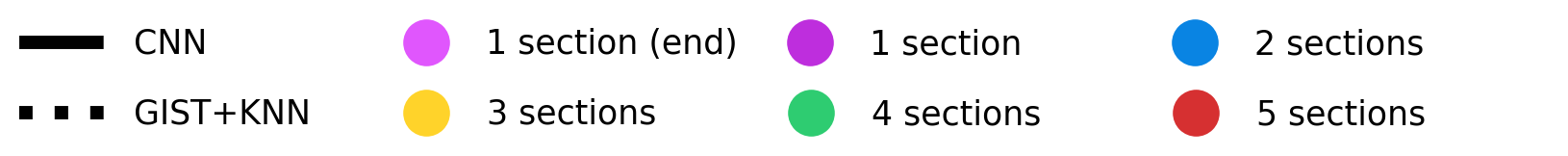}
        \caption{Average accuracy of malware classification under different injection scenarios. Solid lines show results for Inception, and dashed lines for GIST+KNN. Different colors represent the amount of injected sections.}
        \label{inception-results}
    \end{center}
\end{figure}

Some of the challenges involved in building more robust models are closely related to the available data - highly imbalanced number of samples, sample size variation intra and extra families, packing and obfuscation of samples - but those are not the only concerns. Increasing architecture size does not necessarily lead to more robust models, one example being that finetuning a large model like Inception-V1~\cite{Szegedy2016} as done by Chen \textit{et al}~\cite{Chen2018,Chen2020} is also vulnerable to section injection, as seen in a preliminary comparison against KNN+GIST, illustrated in Figure~\ref{inception-results}.

Current results point in the direction of combining text processing techniques with convolutional layers, as made by MalConv~\cite{Raff2017} with its Embedding layer and Le-CNN-BiLSTM~\cite{Le2018} with the recurrent layer after convolutional ones. An open challenge regarding these approaches is related to their input sizes: MalConv truncates data larger than a given size, which requires choosing between discarding relevant data and using more computational resources to process larger samples; CNN-BiLSTM interpolates its input to a fixed size, possibly removing relevant byte relationships in some regions of the file.

\section{Conclusions and Future Work}
\label{sec:conclusion}

In this work we proposed a new method to inject data into malware files in order to change its classification when analyzed by and automatic malware classification system. With a mere 7\% file size increase,  we dropped the accuracy of five classifiers on par with the state-of-the-art - namely GIST+KNN~\cite{Nataraj2011}, MalConv~\cite{Raff2017} and Le-CNN~\cite{Le2018} and two other variations - between 25\% and 40\%. The obtained results seems promising and we think this method can be improved to be robust enough for a larger scale of scenarios. There are some points researchers using this method need to be aware of:

\begin{itemize}

  \item The usage of CNNs is gaining momentum in this research field literature~\cite{Yue2017,Chen2018,Su2018,Khormali2019,Chen2020,Le2018}. This work shows a simple technique that is able to make the accuracy in such CNNs drop in almost 50\% by adding small perturbations to a malware file. We could observe that methods such as Gated CNN~\cite{Raff2017} or combining CNN with LSTM~\cite{Le2018} can be more robust against the data injection presented here.  

  \item A deeper understanding of how the operating system loads executable files to memory usually helps malware creators. During preliminary tests we were able to see that some file format rules are flexible and malware authors do not follow all of them. It includes files with section headers missing or some sections not aligned to the required flags. We tried our best to keep our generated examples in accordance to the format specified. Malware creators might not have this mentality, so that should be considered when building neural networks with the purpose of detecting malware files that rely on static features from the file.

  \item Our results show that data dispersion might be just as important as the amount of data being injected. We can use this idea to conduct a more directed attack using our method together with the method proposed by Khormali \textit{et al.} (2019) \cite{Khormali2019}, injecting FSGM generated sections in any position of the file.

\end{itemize}

As mentioned in Section~\ref{sub:results-defense}, augmenting the training set with injected samples might not be enough to prevent section injection attacks, nor only increasing architecture size. Further investigation is required on how to transform the input for the models in a way that only relevant data for the classification is kept. Current experiments point in the direction that instead of relying on a fixed preprocessing method - like truncating or interpolating - more dynamic approaches should be investigated, such as Attention based methods.

\bibliographystyle{ACM-Reference-Format}
\bibliography{references}

\end{document}